\documentclass[aps,amsmath,amssymb,prb,twocolumn,showpacs,superscriptaddress]
{revtex4}
\usepackage{bm}
\usepackage{epsfig}

\newcommand{\cZ}{{{Z}}}

\newcommand{\br}{{\bf{r}}}
\newcommand{\bN}{{\bf{0}}}

\newcommand{\placefigure}[4]
{
 \begin{figure}[h,floatfix]
 \includegraphics[width=#2]{#1}
 \caption{#3}
 \label{#4}
 \end{figure}
}

\begin{document}

\preprint{draft}

\title{Test of Replica Theory: Thermodynamics of 2D Model Systems\\ with
Quenched Disorder}

\author{Simon Bogner} \affiliation{Institut f\"ur Theoretische Physik,
  Universit\"at zu K\"oln, Z\"ulpicher Stra\ss e 77, 50937 K\"oln,
  Germany} 
\author{Thorsten Emig} \affiliation{Institut f\"ur
  Theoretische Physik, Universit\"at zu K\"oln, Z\"ulpicher Stra\ss e
  77, 50937 K\"oln, Germany} 
\author{Ahmed Taha}
\affiliation{Department of Physics, George Washington University,
  Washington, D.C. 20052, USA} 
\author{Chen Zeng}
\affiliation{Department of Physics, George Washington University,
  Washington, D.C. 20052, USA}

\date{\today}

\begin{abstract}
  We study the statistics of thermodynamic quantities in two related
  systems with quenched disorder: A (1+1)-dimensional planar lattice
  of elastic lines in a random potential and the 2-dimensional random
  bond dimer model.  The first system is examined by a
  replica-symmetric Bethe ansatz (RBA) while the latter is studied
  numerically by a polynomial algorithm which circumvents slow glassy
  dynamics. We establish a mapping of the two models which allows for
  a detailed comparison of RBA predictions and simulations. Over a
  wide range of disorder strength, the effective lattice stiffness and
  cumulants of various thermodynamic quantities in both approaches are
  found to agree excellently. Our comparison provides, for the first
  time, a detailed quantitative confirmation of the replica approach
  and renders the planar line lattice a unique testing ground for
  concepts in random systems.
\end{abstract}


\maketitle

\section{Introduction}

Quenched disorder is common to many condensed matter systems. Examples
include spin glasses \cite{Binder+86}, elastic structures in a random
environment \cite{Nattermann+00} or mesoscopic electronic systems
\cite{Akkermans+94}. In spite of a large volume of theoretical and
experimental work on glasses these systems still pose interesting
challenges. On the theoretical side, much of the studies have focused
on phase diagrams, the existence of phase transitions and their
critical behavior. But even if the equilibrium phases are known,
thermodynamic quantities are in general not accessible.  Slow glassy
equilibration due to the presence of many metastable states on large
length scales is the main obstacle to the numerical and experimental
study of disordered systems. There exist, however, a number of
well-tested analytical tools to study the equilibrium properties.
Common to basically all approaches is that they rely on the
introduction of replicas in order to reestablish translational
invariance. The replicated system is then usually studied by a
renormalization group (RG) approach \cite{DSFisher86} or a Gaussian
variational ansatz (GVA) \cite{Mezard+91} since perturbation theory
completely fails. The first method is designed to yield the effective
pinning potential at large length scales.  The latter approach aims at
constructing a Gaussian trial Hamiltonian which describes the glassy
phase. The results of the RG approaches are, strictly speaking, valid
only close the upper critical dimension or close to a critical point
where randomness becomes irrelevant. On the other hand, the GVA has to
be combined with the concept of "replica symmetry breaking" which is
not generally accepted for elastic structures in random media
\cite{Nattermann+00}. Although the results of both approaches are
similar, the different underlying concepts indicate that the present
general understanding of disordered systems is still incomplete. A
more general qualitative picture was developed in form of the droplet
theory for spin glasses \cite{FisherDS+88}. It provides a
phenomenological scaling approach to static and dynamics properties of
the spin-glass ordered phase. The properties of this phase are
characterized in terms of connected clusters (droplets) of coherently
flipped spins with minimal free energy. It would be desirable to have
a detailed quantitative test of this theory. Progress on numerical
approaches as monte carlo simulations at finite temperatures were
seriously hampered by the slow dynamics. Only recently novel power
full polynomial algorithms became available for the study of large
systems at any temperature \cite{Zeng+96,Zeng+99}.

On the experimental side, most of the effort was devoted to spin
glasses and pinned vortex systems. Especially the latter class
comprises a system which is accessible to exact both analytical
methods and numerical algorithms. It is a randomly pinned planar
vortex lattice that was highlighted by an experimental study of
magnetic flux lines threading through a thin film of the
superconductor 2H-NbSe$_2$ \cite{Bolle+99}. For a certain class of 2D
random systems, including the flux line lattice, new promising
approaches have been developed. For a planar lattice of non-crossing
elastic lines pinned by disorder a replica Bethe ansatz (RBA) can be
employed, yielding exact results for thermodynamic quantities and
their cumulants \cite{Kardar87,Emig+01,Emig+00}. For the related
\cite{Zeng+99} random bond dimer model the partition function can be
calculated exactly by a polynomial algorithm
\cite{Elkies+92a,Elkies+92b} without the need to run slow relaxation
dynamics. Therefore, both method are perfectly suited to overcome to
drawbacks of the above mentioned approaches. However, there are also
limitations to the RBA and the dimer simulations. The first does not
allow to compute correlation functions whereas in the latter the
choice of parameters of the related line lattice model is restricted,
e.g., only one particular density of lines can be simulated. 

The aim of the present work is to show that the thermodynamics of the
two studied model systems provide ideal environments for a
quantitative test of replica theory including replica symmetry
breaking and analytical continuation. Here the 2D model systems can be
considered counterparts of exactly solvable 1D quantum systems which
have advanced the understanding of strongly correlated systems in
general. Our main result is that RBA and dimer model simulations agree
so well that they prove each other to be reliable and thus allow to
explore many question in detail that had been unaccessible to date.

The rest of the paper is organized as follows. In the following
section we introduce the line lattice model and the dimer model, and
explain their mutual relation and their connections to other models.
In section \ref{sec:RBA} we review briefly the replica Bethe ansatz
for the line system and summarize the results in the known limiting
cases. We continue in section \ref{thermodynamics} with a detailed
quantitative comparison of the Bethe ansatz predictions for various
thermodynamic quantities and the corresponding simulation data for the
dimer model. To do so, the Bethe ansatz equations are solved
numerically outside the validity range of the previously studied
limiting cases. We close with a summary and discussion of our results
in section \ref{sec:discussion}.

\section{The models and their Connections}
\label{sec-models}

\subsection{Vortex system}
\label{vortex_system}

We consider an ensemble of directed vortex lines confined to a plane
at average distance $a\equiv1/\rho$. The configurations of a single
directed line is characterized by its position $x_i(z)$ since
overhangs are forbidden. The prefered path of the line results from
the competition between elastic energy, measured by the line tension
$g$, the line interaction in form of a repulsive pair potential $U(x)$
that does not allow the lines to cross and pinning by a random
impurity potential.  With a contact repulsion $U(x)=c\;\delta(x),
c\to\infty$, ensuring the noncrossing condition, the model remains
generic\cite{Emig+03} and allows for the mapping to a discrete dimer
model. Quenched disorder couples locally to the vortices via a random
potential $V(\br)$, which we assume to have zero mean and short-range
correlations
\begin{equation}
\label{eq:VV-corr}
\overline{V(\br)V(\br')}=\Delta\delta_{\xi_d}(\br-\br').
\end{equation}
Whenever the disorder correlation length $\xi_d$ is the smallest scale
in the problem it can safely be set to zero. The total energy
can be written as
\begin{equation}
\label{line_hamiltonian}
H=\int dz \sum_i \left\{\frac{g}{2}\left(\frac{dx_i}{dz} \right)^2 +
c\sum_{i,j\ne i} \delta(x_i-x_j)+V(x_i,z) \right\}.
\end{equation}
Throughout the paper, the disorder average will be denoted by
$\overline{\phantom{I}\dots\phantom{I}}$ and the thermal average by
$\langle\phantom{I}\dots\phantom{I}\rangle$.

\subsection{Dimer model}
\label{sec:dimer_model}

\begin{figure}[h]
\includegraphics[width=1\linewidth]{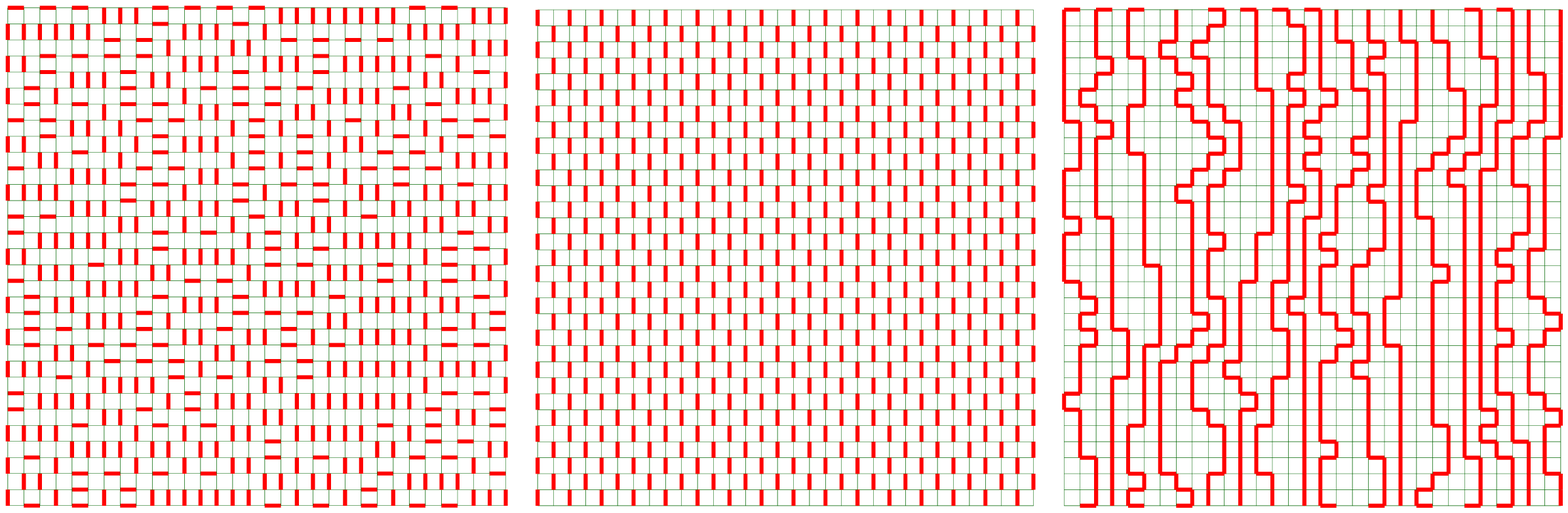}
 \caption{Mapping of a dimer
  (left) to a line configuration (right) via XOR-addition of the reference
  state (middle).}
 \label{mapping-large}
\end{figure}

The dimer model is defined as follows: Choose a subset (whose elements
are called dimers) of the bonds on a square lattice with lattice
constant $b$ and linear size $L$ such that every of the $L^2=N$
lattice sites (labelled by $(ij)$) is touched by exactly one of these
dimers, see Figs.~\ref{mapping-large},~\ref{dimers}. A square lattice
rotated by 45 degrees with lattice constant $b/\sqrt2$ is formed by the
centers of the bonds.  Its $2N$ sites shall for convenience also be
labelled by $(ij)$ and it will be clear from the context if the
original lattice or that of the bonds is parametrized. The reduced
energy of one such complete covering $D$ of $N/2$ dimers is
defined by
\begin{equation}
\label{H-dimer}
H_{d}=\sum_{(ij)\in D} \epsilon_{ij}/T_d, 
\end{equation}
where the sum is over all dimers of $D$. The bond energies
$\epsilon_{ij}$ are randomly drawn from a Gaussian distribution with
zero mean and unit variance
$\overline{\epsilon_{ij}\epsilon_{kl}}=\delta_{(ij),(kl)}$.  $T_d$ is
the dimer temperature and measures the strength of disorder.  The
implementation of the polynomial algorithm
\cite{Elkies+92a,Elkies+92b} (with exponent $\simeq 2$) on a
32-processor cluster allows to compute thermodynamic quantities
numerically exactly -- as opposed to, e.g., Monte Carlo sampling --
for sizes up to $L=512$ at typically $6000$ disorder configurations
within a CPU time of days.  Merely the measurement of the specific
heat in Section \ref{response_functions} is not covered by the
polynomial algorithm.  Thermal fluctuations have to be computed from
explicitly sampling over a representative set out of $\exp(N G/\pi)$
possible dimer coverings \cite{Kasteleyn61}, again for up to 6000
disorder configurations. This gives, however, also reliable results
for sizes up to $L=256$.  The typical accuracy of numerical data at
the given number of disorder samples is $\simeq 10^{-5}$. For further
details of the algorithm for the simulation of the dimer model see,
e.g., Ref.~\onlinecite{Zeng+96}.

\subsection{Connections}
\label{sec_mapping}

These two apparently diverse models are closely related.  The random
bond dimer model can be mapped onto an array of lines that interact
via a hard-core repulsion, preventing any line crossing.  In
Fig.~\ref{mapping} this mapping and the connections to related models
are sketched.  From the dimer model (A) on top, the discrete lattice
version of the line model (C) can be reached directly or via the
intermediate random solid-on-solid (SOS) model (B). Both the SOS model
and the discrete lines have their continuous counterparts -- in the
bottom line of the sketch -- that can be treated analytically: The
2-dimensional random sine-Gordon (RSG) model (D) (being equivalent to
the random-field XY model without vortices) and the continuum
(1+1)-dimensional elastic lattice of directed lines (E).  The
parameters and observables of the isotropic dimer model and its
version with a height profile (SOS) can be expressed in terms of the
anisotropic vortex line lattice.  In the following we describe the
relations between the above mentioned models in detail.  We start the
tour through the mapping table of Fig.~\ref{mapping} from the dimer
model (A) with the dimer temperature $T_d$ as the only control
parameter. The discrete line lattice can be reached easily via the map

{\em (AC) -- } Add a given dimer pattern and a regular reference dimer
covering as given in the middle of
Figs.~\ref{mapping-large},~\ref{dimers} with an "exclusive or" (XOR)
operation. Only if a given bond is covered by either the dimer pattern
or the reference pattern it shall be covered in the resulting line
configuration, which will be noncrossing lines at average density
$\rho=1/(2b)$. A resulting line lattice configuration is shown in the
right part of Figs.~\ref{mapping-large}, \ref{dimers}.  For the summed
energies of all covered bonds in the respective configurations it
holds
\begin{equation}
\label{energy_relation}
      H_{\rm l}(\{\epsilon'_{ij} \})=H_d(\{\epsilon_{ij}\})+
      H_{\rm ref}(\{\epsilon'_{ij} \}).
\end{equation}
Here and in the following, the subscript '$d$' stands for quantities
of the dimer model, while line-lattice quantities are denoted by the
subscript '$l$'.  $\{\epsilon_{ij}\}$ stands for a given distribution
of random energies on all of the bonds of the dimer model while the
set of random energies $\{\epsilon'_{ij}\}$ are defined as
$\epsilon'_{ij}=-\epsilon_{ij}$ on the occupied bonds of the reference
pattern and $\epsilon'_{ij}=\epsilon_{ij}$ elsewhere. If the original
random bond energies $\{\epsilon_{ij}\}$ are distributed symmetrically
with zero mean, so are the bond energies $\{\epsilon'_{ij}\}$ defining
the discrete line lattice model.

\placefigure{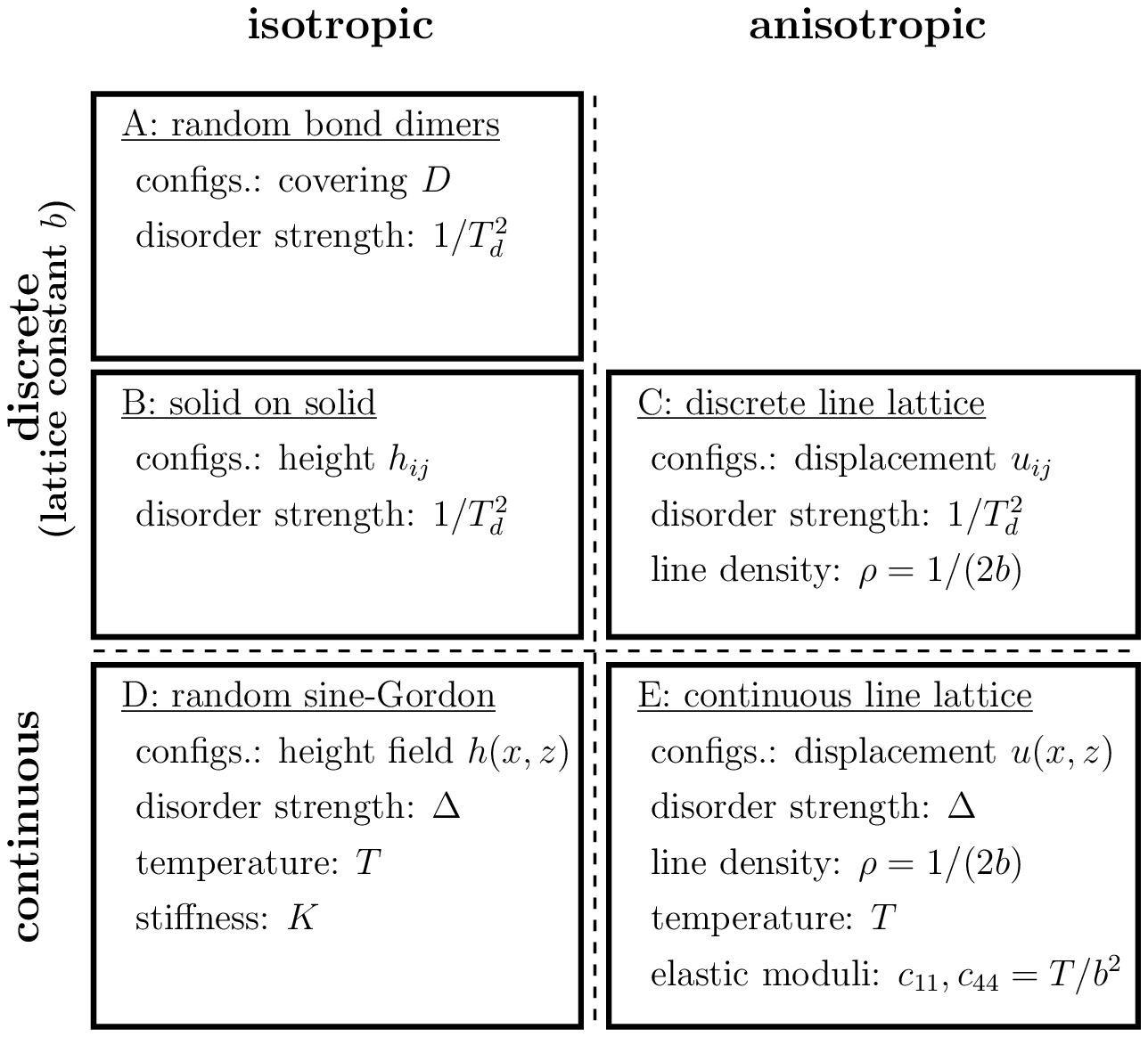}{0.98\linewidth}{Overview of the models introduced
  in the text, showing their degrees of freedom and parameters.  {\em Top}:
  discrete models; {\em bottom}: continuous models ; {\em left}:
  isotropic models, {\em right}: anisotropic models.}{mapping}

{\em (AB) -- } A discrete height profile $\{h_{ij}\}$ can be assigned
to every plaquette of the square lattice in the following way
\cite{Henley97}: Every bond is given a sign $\pm 1$ such that when
going through rows or columns of plaquettes the sign of the crossed
bonds alternates.  Starting at a given plaquette with arbitrary
height, one moves to the neighboring plaquettes and adds to the height
$+3$ times the bond sign if the crossed bond is covered by a dimer, or
$-1$ times the bond sign if it is not. The resulting numbers define a
path-independent height profile $\{h_{ij}\}$, see Fig.~\ref{dimers}
for an example.

\placefigure{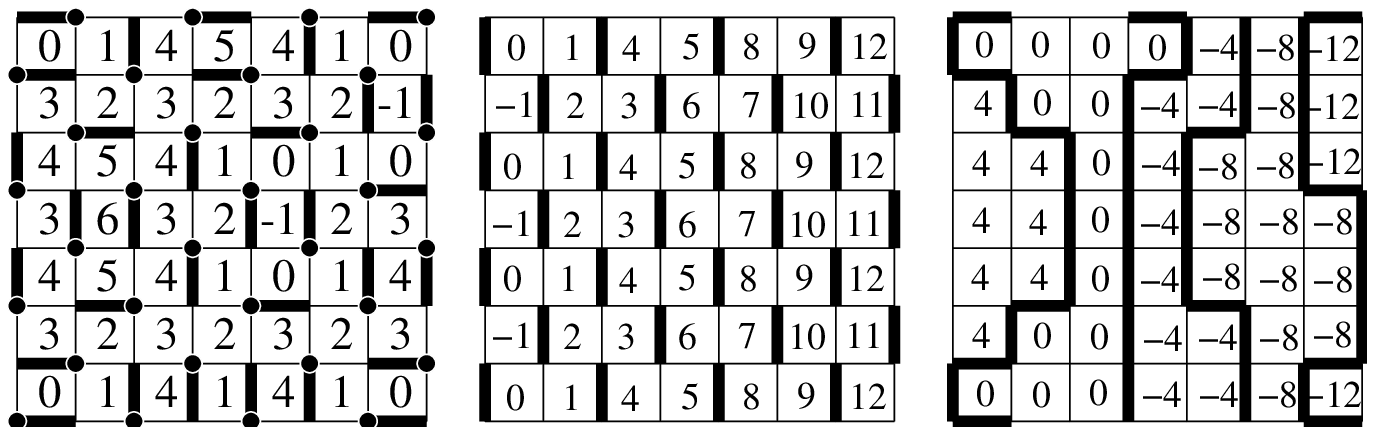}{0.95\linewidth}{Dimer covering (left), reference
  state (middle) and line configuration (right), with height profiles
  $h_{ij}$ (left), $h_{ij}^\text{ref}$ (middle) and $H_{ij}$
  (right).}{dimers}

{\em (BC) -- } Define a new discrete height profile $\{ H_{ij}\}$ on
each plaquette by subtracting the height profile $\{
h_{ij}^\text{ref} \}$ associated with the reference dimer covering,
cf. the middle part of Fig.~\ref{dimers},
\begin{equation}
\label{H}
H_{ij} \equiv h_{ij}-h^\text{ref}_{ij}.
\end{equation}
The profile $\{H_{ij}\}$ is then quantized in steps of width 4.
Plaquettes of constant $H$ form domains of average width $2b$ that may
be numbered by $k$. The domains are separated by domain walls which
can be considered as directed lines.  Next, we introduce the
displacements $u_k(i)$ of these domain walls from their perfectly
aligned positions where $i=1,\ldots , L$ and $j=2k$. As illustrated in
Fig.~\ref{profiles}, the line displacements are determined by the
height profile, leading to
\begin{equation}
\label{u-H}
u_k(i)=\frac{H_{i,2k}+H_{i,2k+1}}{4}+2k.
\end{equation}
It is important to keep in mind that the mapping to a displacement
field makes sense only for a well-defined initial configuration of the
domain walls implying a {\em fixed} density. In the grand canonical
ensemble induced by the sum over dimer configurations, $u_k(i)$ is a
good degree of freedom only for the configurations corresponding to
the mean line density $\rho=1/(2b)$, which, however, carry the
dominant weight in the thermodynamic limit where the density
distribution is sharply peaked.

\placefigure{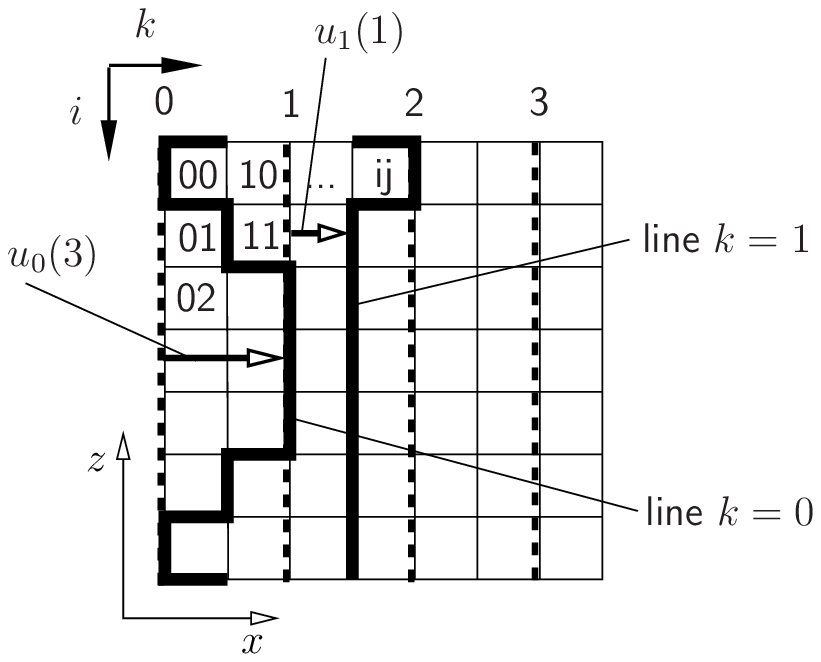}{0.7\linewidth}{Illustration of
  displacements $u_k(i)$ in the discrete model.}{profiles}

{\em (CE) -- } In a continuum formulation each individual line can be
considered as a directed polymer with energy $\frac{g}{2}\int dz
\left\{\partial_z x_i(z)\right\}^2$ and line tension $g$. If one
includes the non-crossing condition imposed by the underlying dimer
configurations and the coupling to disorder the continuum model is
that introduced in Eq.~(\ref{line_hamiltonian}). In a second step this
system can be cast into the form of a 2-dimensional elastic system. We
define a continuous displacement field $u(\br)$ so that the line
positions are $x_i(z)=i a +u(ia,z)$. If we include the line
interaction in the elastic energy, the Hamiltonian for fluctuations on
sufficiently large length scales can be written as
\begin{equation}
\label{cont_lines}
H_{\rm el}=\int\! d^2\!\br \left\{\frac{c_{11}}{2}(\partial_x u)^2 +
\frac{c_{44}}{2}(\partial_z u)^2
+\rho(\br)V(\br)\right\}
\end{equation}
with compression modulus $c_{11}=a U''(a)$, tilt modulus $c_{44}=g/a$
and line density $\rho(\br)=\sum_j \delta(x-x_j(z))$. The correlations
of the random potential $V(\br)$ are given by Eq.~(\ref{eq:VV-corr}).
The compression modulus is obtained in elastic approximation from the
line interaction potential $U(x)$,
\begin{equation*}
c_{11}=a\frac{\partial^2}{\partial a^2} U(a).
\end{equation*}
For the special contact repulsion $U(x)=c \delta(x)$, $c\to\infty$,
imposed by the equivalence to the dimer model, the microscopic (short
scale) $c_{11}$ vanishes. However, on larger scales (or for more
general microscopic interactions) $c_{11}$ assumes a finite value due
to entropic contributions. The resulting macroscopic $c_{11}$ can be
obtained from the free energy density $f$ of the model of
Eq.~(\ref{line_hamiltonian}) by \cite{Landau+69}
\begin{equation}
\label{compression_modulus-definition}
c_{11}=a\frac{\partial^2}{\partial a^2}\left[ a f(a)\right].
\end{equation}

{\em (BD) -- } The continuum limit of the random SOS model is the
random sine-Gordon RSG model with the reduced Hamiltonian
\cite{Zeng+99}
\begin{equation}
\label{H_RSG}
H_{\rm \tiny RSG} = \int d^2 \br \left\{ \frac{K}{2} (\nabla h)^2 
+ \frac{V'}{T} \cos[Q h(\br)+\alpha(\br)]\right\},
\end{equation}
where $K$ is the reduced surface stiffness, $\alpha(\br)$ is a
spatially uncorrelated and uniformly distributed random phase, $V'$
the coupling strength of disorder and $Q=2\pi/4$.  The proper
periodicity $Q$ of the disorder coupling can be understood from the
periodicity of the related discrete line lattice, i.e., via the path
ACED in the diagram of Fig.~\ref{mapping}. The latter sequence of
mappings also allows to conclude about the value of the stiffness $K$
in terms of the parameters of the discrete model, see below.

{\em (ED) -- } The two continuum models D and E differ by their form
of elasticity (isotropic vs. anisotropic) and the disorder coupling.
If we set $h(\br)=4u(\br)/a$ and rescale the $z$-axis in model E
according to $z\rightarrow z'=z\sqrt{c_{44}/c_{11}}$ we obtain the
isotropic elasticity of model D, cf.~Eq.~(\ref{H_RSG}), with the
stiffness given by
\begin{equation}
\label{K}
K=\frac{a^2\sqrt{c_{11}c_{44}}}{16T}.
\end{equation}
The disorder energy of Eq.~(\ref{cont_lines}) can be transformed to
the form of Eq.~(\ref{H_RSG}) by expanding the local density of the
lines $\rho(\br)$ in terms of the line displacement $u(\br)$ by use
of Poisson's summation formula, for details see Ref.
\onlinecite{Nattermann+00}.  The relation $h(\br)=4u(\br)/a$ between
the line displacement field and the (dimensionless) height profile is
the continuum version of Eq.~(\ref{u-H}) which can be seen with the
use of Eq.~(\ref{H}) and the observation that $h_{ij}^{\rm ref}\simeq
2j$ upon coarse graining.  This relation is in hindsight, knowing that
the lines have a mean distance of $a=2b$, the reason for the choice of
$Q=2\pi/4$ in model D, Eq.~(\ref{H_RSG}). Due to the coordinate
rescaling, the system size $A=L_x L_z$ is changed when going from
model E to D. If we denote the system sizes by $A_\text{el}$ and
$A_\text{RSG}$, respectively, we obtain the ratio
\begin{equation}
\label{volume_relation}
\frac{A_\text{RSG}}{A_\text{el}}=\sqrt{\frac{c_{11}}{c_{44}}}=K \frac{8T}{bg}.
\end{equation}

Having explained the relations between the models, we have to specify
how the three continuum model parameters, (i) line stiffness $g=a \,
c_{44}$, (ii) disorder strength $\Delta$ defined by
Eq.~(\ref{eq:VV-corr}) and (iii) temperature $T$, are related to the
dimer model.  First, we observe that only the relative strength of the
parameters $g$ and $\Delta$ with respect to thermal fluctuations is
important, i.e., the ratios $g/T$ and $\Delta/T^2$ have to be
determined.  Out of these we start with the reduced single line
stiffness $g/T$.  In the dimer model, out of the $2L^2$ bonds on
average likewise $L^2/4$ horizontal and vertical bonds are occupied.
This holds both with respect to thermal sampling, i.e., summing over
dimer configurations, and the disorder average as long as the mean
random energy is isotropic.  For the discrete lines after the "XOR"
addition with the reference state, an average number of $L^2/4$
horizontal and $L^2/2$ vertical segments is implied, corresponding to
a mean line density of $\rho=1/(2b)$.  We now consider a given
discrete line as performing a one-dimensional random walk with the
three possibilities of moving to the left or to the right, or to rest.
The probabilities for the possible steps $x_i=-1,0,1$ can be deduced
from the average number of occupied horizontal and vertical bonds.
From the above analysis of occupied bonds we find the corresponding
probabilities $w_i=1/4,1/2,1/4$ for the steps $x_i$.  For the
fluctuations of the total horizontal wandering $X\equiv\sum_i x_i$
this leads after $z/b$ steps to
\[
\langle X^2 \rangle=\sum_{i=1}^{z/b}\langle x_i^2\rangle= z\; b/2.
\]
A continuum Hamiltonian for a random walk $H=\frac{g}{2}\int dz
(\partial_z X)^2$ in comparison yields $\langle X^2
\rangle=\frac{T}{g} z$ and allows to read off
\begin{equation}
\frac{gb}{T}=2.
\end{equation}
\begin{table*}
\begin{tabular}{lp{.9cm}lp{.9cm}lp{0.9cm}lp{.9cm}l}
\hline
model && line density && line tension && disorder && system size\\
\hline\hline
continuum line lattice&&$\rho\equiv 1/a$ &&$g/T$ &&$\Delta/T^2$&& 
$L_x \times L_z$\\\hline
isotropic dimer model && $1/(2b)$&& $2/b$ && $\simeq 2\; \xi_d/(bT_d^2)$&& 
$bL \times \frac{ 16 K T}{ag} bL$\\\hline
anisotropic dimer model  && { "} && { ''} && 
$\simeq \xi_d/(bT_d^2)$ && { "}\\\hline
\end{tabular}
\caption{Relation between parameters of the continuum line lattice and 
the dimer model with random vertical  {\em and} horizontal bond 
energies (isotropic) or only random vertical bond energies (anisotropic).}
\label{mapping_table}
\end{table*}
The strength of disorder is measured in the continuum theory by the
variance $\Delta$ of the disorder potential. Noting that $1/T_d^2$ is
the variance of the (reduced) random bond energies in the dimer model
we are lead to identify $\sqrt{\Delta}/T$ with the dimensionless
inverse dimer temperature $1/T_d$.  Allowing for the finite disorder
correlation length $\xi_d$, cf.~Eq.~(\ref{eq:VV-corr}), which acts as
a cutoff in the continuum model, to differ slightly from the lattice
constant $b$ of the dimer model, we have
\begin{equation}
\label{first_guess}
\frac{\Delta}{T^2}=\frac{\xi_d}{b}\frac{1}{T_d^2}.
\end{equation}
A closer look at the models, however, suggests that this relation is
valid only if the energies $\epsilon_{ij}$ of the horizontal bonds of
the dimer model are set to zero, i.e., if there is no disorder on
these bonds.  This is because in the continuum model of
Eq.~(\ref{cont_lines}) the disorder energy is written as the coupling
of the local line density $\rho(\br)$ to the disorder potential
$V(\br)$,
\begin{equation*}
  H_{\rm dis}=\int\!\! dx dz\, V(x,z) \rho(x,z)=\sum_i \int\!\! dz\, V(x_i(z),z).
\end{equation*}
The random part of the energy is thus not an integral over the arc
length of the lines but over their $z$-coordinate.  Therefore, the
disorder energy of a line is proportional to its length projected onto
the $z$-axis and not to its overall length.  In the original isotropic
dimer model, random energy is collected on both vertical and
horizontal bonds and will thus be proportional to the overall length
of a fluctuating line.  As a consequence, a given disorder in the
dimer model corresponds to larger disorder in the continuum line model
than assumed by Eq.~(\ref{first_guess}). A possible method to gauge
the disorder strength it to consider the annealed disorder average
$-T\ln\overline Z$ of the free energy. This quantity can be obtained
analytically in both models.  The calculation for the continuum model
is straightforward whereas in the dimer model the means of
Ref.~[\onlinecite{Kasteleyn61}] together with the relation of
Eq.~(\ref{energy_relation}) can be used. One finds by comparison
between the two models the relation
\begin{equation}
\label{disorder_via_annealed}
\frac{\Delta}{T^2}=\frac{\xi_d}{b}\frac{2}{T_d^2}\Big\{1- \frac{2GT_d^2}
{\pi}+\frac{2T_d^2}{\pi}\int_0^{e^{-1/(2T_d^2)}}\!\!\!dx\,
\frac{\arctan x}{x}\Big\}
\end{equation}
with Catalan's constant $G=0.915966$.  The contribution from the
latter two terms in the curly brackets crosses over at $T_d\simeq 1$
from zero at small $T_d$ to $-1/4$ at large $T_d$. Thus the overall
factor of two as compared to the naive estimate of
Eq.~(\ref{first_guess}) confirms the conjecture about the contribution
of the horizontal random bond energies for strong disorder (small
$T_d$). For weak disorder (large $T_d$) the gauging by the annealed
free energy yields a factor of $3/2$ compared to
Eq.~(\ref{first_guess}).  However, in most expressions below, the
latter two terms will only act as corrections at intermediate $T_d
\simeq 1$ and are in those cases frequently neglected.  (At large
$T_d$ constant additional terms which are not related to the gauging
procedure dominate.) It should be noted that the exact relation
Eq.~(\ref{disorder_via_annealed}) has been derived for one special
observable, the annealed free energy, and it cannot be expected to
hold universally for all observables with the same value for the
regularization length $\xi_d$.  Rather, to different observables the
short scale modes around the UV cutoff can contribute with a different
weight.  The one free parameter $\xi_d/b$ relating the continuum model
to the discrete model will therefore be considered a fitting
parameter, which should, however, not turn out to vary dramatically
around its expected value of order unity from observable to
observable.

Below, results from numerical simulations of both the isotropic dimer
model with random energies on vertical and horizontal bonds and the
dimer model with vanishing energies on the horizontal bonds will be
compared to theory.  For the former, Eq.~(\ref{disorder_via_annealed})
will be used as disorder strength mapping (sometimes without the
correction terms) while for the latter Eq.~(\ref{first_guess}) will
prove to fit very well.  Table \ref{mapping_table} summarizes the
relations between the dimer parameters and the continuum model
parameters.

\section{Replica Bethe ansatz}
\label{sec:RBA}

The classical statistics of fluctuating elastic lines in
$d$ dimensions can be described by the quantum statistics of a
$(d-1)$-dimensional system of interacting bosons \cite{Nelson+89}. In
the thermodynamic limit, the free energy and its disorder fluctuations
are determined by the ground state energy of the Bose gas. An
important simplification arises for $(1+1)$-dimensional systems of
self-avoiding lines for two reasons. First, in the absence of quenched
disorder the interacting Bose gas is replaced by a $1$-dimensional
free Fermi gas since the non-crossing condition for the lines is then
automatically fulfilled by the Pauli exclusion principle
\cite{DeGennes68,Pokrovsky+79}. Second, $1$-dimensional quantum
systems with sufficiently simple interactions (generated here by
quenched disorder) can be often treated exactly by Bethe ansatz.
Indeed, Kardar used the replica method to show that the self-avoiding
line lattice in a random potential maps to a gas of fermions with $n$
spin components with $SU(n)$ symmetry interacting via an attractive
$\delta$-function potential \cite{Kardar87}. The ground state energy
of this system can be calculated exactly by Bethe ansatz.  The analogy
between the replicated line lattice and $SU(n)$ fermions was examined
further in Ref.~\onlinecite{Emig+01}.  In the following we will
summarize the main results of the replica Bethe ansatz (RBA).

Upon replication of the system of Eq.~(\ref{line_hamiltonian}) and
disorder averaging with the aid of Eq.~(\ref{eq:VV-corr}), the
equivalent quantum system is described by the Hamiltonian
\begin{equation}
  \label{eq:QM-hamiltonian}
  \hat H = -\frac{T^2}{2g} \sum_{\alpha=1}^n \sum_{j=1}^N
  \frac{\partial^2}{\partial x_{j,\alpha}^2} - \frac{\Delta}{T} 
  \sum_{\alpha<\beta} \sum_{j,k} \delta(x_{j,\alpha}-x_{k,\beta}),
\end{equation}
where $N$ is the number of lines and $n$ the number of replicas.  In
the quantum system the line stiffness $g$ corresponds to the fermion
mass, the temperature $T$ is mapped onto $\hbar$ and the system size
in $z$-direction, $L_z$, onto $\hbar\beta$ where $\beta$ is the
inverse quantum temperature.  We are interested in the
ground state with $SU(n)$ symmetry and the corresonding energy
$E_0(n)$.  As a function of the replica index $n$, the ground state
energy $E_0$ carries information on the statistics (cumulants) of the
line lattice free energy.  The disorder averaged moments of the line
lattice partition function are given by
\begin{equation}
\label{Z_replica}
\ln \overline{\cZ^n}=\sum_{j=1}^\infty \frac{(-n)^j}{j!} 
\frac{\overline{F^j}_c}{T^j}=-\frac{E_0(n)L_z}{T}+n\frac{L_x}{a} \ln \cZ_0.
\end{equation}
Here, $\overline{F^j}_c$ are the cumulants of the disorder distributed
line lattice free energy and $\cZ_0$ is the partition function of a
single line with no disorder contributions. The above expansion in the
number of replicas $n$ relies on the assumption of analyticity of the
replica free energy (or ground state energy) around $n=0$. That this
assumption is indeed justified will be demonstrated below.

The model of Eq.~(\ref{eq:QM-hamiltonian}) is integrable and its
ground state energy can be obtained in terms of $n$ nested Bethe
ans\"atze as demonstrated by Sutherland \cite{Sutherland} for the
repulsive case, and by Takahashi \cite{Takahashi70} and Kardar
\cite{Kardar87} for the attractive case of interest here. For the
Bethe ansatz calculation $n$ is assumed an integer number. In order to
extract information from the ground state energy about the
thermodynamics of the line lattice, $E_0(n)$ has to be analytically
continued to real valued $n$. $E_0(n)$ in the limit $n\to 0$ has
been obtained in Ref.~\onlinecite{Kardar87}.  Using a different method
for the analytical continuation, an expression for $E_0(n)$ at
arbitrary $n$ was derived in Ref.~\onlinecite{Emig+01}. The result of
the latter reference can be summarized as follows. The Bethe ansatz
equations can be analytically continued and thus yield the allowed
wave numbers of the ground state wave function for general $n$.  The
ground state energy can then be expressed in terms of the density
function $\varrho(k)$ which yields the number $L_x \varrho(k)dk$ of
allowed wave numbers in the interval $[k,k+dk]$. One finds
\begin{equation}
\label{E_0}
E_0(n)=\frac{g\rho\Delta^2 L_x}{24 T^4}n(1-n^2)+n \frac{T^2 L_x}{2g}
\int_{-K}^K \!dk\; k^2 \varrho(k).
\end{equation}
The ($n$ dependent) density function $\varrho(k)$ is determined by the
integral equation
\begin{subequations}
\begin{equation}
\label{BA_equation1}
nk=\int_{-K}^K\!dk' g_n\left[(k-k')l_d\right]\varrho(k') 
\end{equation}
with the kernel
\begin{equation}
\label{kernel}
g_n(x)=2\sum_{m=0}^\infty \arctan\left[\frac{nx}{m^2+nm+x^2}\right].
\end{equation}
The length $l_d=T^3/(g\Delta)$ is the characteristic length scale of
the interaction in the quantum problem. This length sets the crossover
scale beyond which the line fluctuations are dominated by disorder.
In the above equations, the integral boundary $K$ is fixed
by the mean line density via
\begin{equation}
\label{BA_equation2}
\rho=\int_{-K}^K \!dk \; \varrho(k).
\end{equation}
\end{subequations}
In the limit $n\to 0$, the integral equation (\ref{BA_equation1})
assumes the form \cite{Kardar87,Emig+01}
\begin{equation}
  \label{eq:BA_eq_n0}
  \int_{-K}^K dk' \left\{ \frac{1}{l_d(k-k')} + 
    \pi \coth\left(\pi l_d(k-k')\right) \right\}\varrho(k') =k.
\end{equation}
This equation can be solved perturbatively in $K l_d$, yielding
\begin{eqnarray}
  \label{eq:rho-sol-n0}
  \varrho(k)&=&\sqrt{1-(k/K)^2}\left[ \frac{1}{2\pi} K l_d -
\frac{\pi}{24} (K l_d)^3 \right. \nonumber \\
&& \left. + \frac{\pi^3}{144}\left(1+\frac{2}{5} (k l_d)^2
\right) (K l_d)^5 + \ldots \right].
\end{eqnarray}
In the opposite limit of vanishing disorder or $l_d \to \infty$, the
exact solution is $\varrho(k)=1/(2\pi)$. Using Eqs.~(\ref{E_0}),
(\ref{BA_equation2}) the ground state energy can be calculated from
Eq.~(\ref{eq:rho-sol-n0}) perturbatively in $K l_d$. Interestingly, it
can be shown order by order in $K l_d$ that the {\em exact} result for
the ground state energy is obtained from the first two terms in the
square brackets of Eq.~(\ref{eq:rho-sol-n0}). Indeed one finds
that \cite{Emig+03} 
\begin{equation}
\label{eq:kinetic-n0}
\int_{-K}^K dk k^2 \varrho(k) = \frac{\pi^2}{3} \rho^3 + \rho^2
l_d^{-1}
\end{equation}
to any order in $\rho l_d$. Using $\overline{f}=\lim_{n\to 0}
E_0(n)/(nL_x)$ this leads to the mean free energy density of the line
system,
\begin{equation}
\label{line_free_energy}
\overline{f}=\overline{f_0}\rho+\frac{\pi^2}{6}\frac{T^2}{g}\rho^3
+\frac{\Delta}{2T}\rho^2,
\end{equation}
where the contribution from single non-interacting lines is given by
\begin{equation}
\label{single_line_f}
\overline{f_0}=C(g)T-\frac{\Delta}{2\xi_d T}+
\frac{g\Delta^2}{24T^4},
\end{equation}
with $C(g)$ a disorder independent contribution that cannot be
obtained from Bethe ansatz.  The result for interacting lines
[Eq.~(\ref{line_free_energy})] was first derived in
Ref.~\onlinecite{Emig+03} while the latter result for a single line
was first given in Refs.~[\onlinecite{Kardar+85,Kardar87}] with finite
size corrections calculated in
Refs.~[\onlinecite{Bouchaud+90,Brunet+00}].  It should be noted that
in the interaction energy of Eq.~(\ref{line_free_energy}) the entropic
thermal contribution -- second term -- and steric disorder
contribution -- third term -- are simply additive with no interference
between the two effects.  They agree in their dependency on the system
parameters with the respective expressions based on scaling
arguments.\cite{Pokrovsky+79,Kardar+85}

So far we have only used the integral equation (\ref{BA_equation1}) in
the limit $n \to 0$. However, this equation taken at small $n$
contains information about the cumulants of the free energy.
Therefore, we would like to compute the ground state energy $E_0(n)$
perturbatively in $n \ll 1$. In general, this is not feasible by
analytical means. Therefore, below we will solve
Eq.~(\ref{BA_equation1}) numerically in order to extract the behavior
of $E_0$ for small $n$. There is, however, one important limiting case
where analytical progress is possible. For strong disorder or low line
density with $l_d \ll 1/\rho$, the integral Eq.~(\ref{BA_equation1})
has been shown to assume a particular interesting form.  It reduces
\cite{Emig+01} after a rescaling to the Bethe ansatz equation for the
$1$-dimensional Bose gas with a {\em repulsive} $\delta$-function
interaction \cite{Lieb+63}. The effective interaction strength of the
Bose gas is proportional to $n^2/l_d$. Therefore, in the interesting
limit of small $n$ one can either use the numerical solution of the
Bethe ansatz equations for the Bose gas \cite{Emig+01} or Bogoliubov's
perturbation theory \cite{Bogoliubov47} to calculate the ground state
energy $E_0(n)$ of the $SU(n)$ fermions as a polynomial in $n$ to
lowest order in $\rho l_d$. From $E_0(n)$ the quenched averaged
moments of the partition function of the line system can be evaluated
using Eq.~(\ref{Z_replica}). For the disorder dependent contributions
one finds the result \cite{Emig+01}
\begin{eqnarray}
\label{moments_perturbative}
&&\ln \overline{\cZ^n}
=-\frac{L_x L_z}{2}\Big(\frac{\Delta}{T^2}\Big)^3\Big(\frac{g}{T}\Big)^2
\bigg\{ \frac{n}{12}(1-n^2)l_d\rho+n(l_d\rho)^2 \nonumber\\
&&-\frac{4}{3\pi}n^2(l_d\rho)^{3/2}+\left(\frac{1}{6}-\frac{1}{\pi^2}
\right)n^3 l_d\rho + {\cal O}\left(n^4\right)   \bigg\}.
\end{eqnarray}
The first term $\sim (1-n^2)\rho$ in the curly brackets describes the
disorder contribution to the single-line free energy cumulants while
the following terms stem from line interactions in the presence of
disorder. As can be seen from Eq.~(\ref{Z_replica}) the above result
provides the free energy cumulants in the dilute limit $\rho l_d \ll
1$. At higher densities, the mapping to the Bose gas is no longer
valid and the ground state energy has to calculated by direct
numerical solution of the integral Eq.~(\ref{BA_equation1}). This will
be done in Sec.~\ref{higher_moments}.

A natural limit to the validity of the RBA results is set by the
mapping of the line system with short-ranged correlated disorder to
the quantum problem with a singular $\delta$-function interaction
arising from the disorder average.  The characteristic length scale in
the quantum problem is $l_d$ which is inversely proportional to the
disorder strength $\Delta$.  If $l_d$ becomes of the order of the
cutoff length $\xi_d$ of the short-ranged disorder correlator of
Eq.~(\ref{eq:VV-corr}), the assumption of ultra-locally interacting
fermions is no longer justified.  The $\delta$-function interaction,
however, is essential for the Fermi gas to be solvable by Bethe
ansatz.  The RBA results therefore are valid only for temperatures
\begin{equation}
\label{T_star}
T\gtrsim T^*=(g \xi_d \Delta)^{1/3},
\end{equation}
or, respectively, at sufficiently weak disorder.  For lower
temperatures a modified replica symmetry breaking solution has been
suggested in Ref.~[\onlinecite{Korshunov+98}] for the single line, the
predictions of which, however, could not be tested to date by other
means. For the interacting line system the modifications at low
temperatures implied by the replica symmetry breaking solution can be
adapted.\cite{Emig+01} However, when the RBA results are translated to
the random bond dimer model it turns out that the low temperature
limit $T<T^*$ can be never realized in the latter model.

\section{Thermodynamics}
\label{thermodynamics}

\subsection{Large scale equivalence}

Before studying thermodynamics, we will demonstrate that the
statistics of the random bond dimer model and its associated discrete
height profile on large length scales are well described by the
continuum model for elastic lines which can be treated analytically by
the RBA.  From the simulation of the random bond dimer model, the
correlations of the height profile $\{h_{ij}\}$ of model B can be
determined very accurately.  The quantity $\delta
h(\br)=h(\br)-\langle h(\br)\rangle$ measures the thermal fluctuations
of the height around its state pinned by disorder.  A Renormalization
group calculation predicts for the disorder averaged correlation
function\cite{Nattermann+00}
\begin{equation}
\label{eq:hh-corr}
C(\br)=\overline{\langle [\delta h(\br)-\delta h(\bN)]^2\rangle}=
\frac{1}{\pi K}\ln(|\br|/b),
\end{equation}
i.e., logarithmic growth on large scales, irrespective of the value of
$\langle h(\bN)\rangle$. A possible transition from a glassy low
temperature phase with $\langle h(\bN)\rangle>0$ to a free thermal
phase with $\langle h(\bN)\rangle=0$ therefore is not reflected
directly in this correlation function.  However, the coefficient
$1/(\pi K)$ will be of interest.  The stiffness $K$ obtained from a
measurement of the correlation function for large $|\br|$ is the {\em
  large-scale} effective stiffness, renormalized by contributions from
thermal and disorder fluctuations on smaller scales. It can be
calculated exactly from the RBA free energy
Eq.~(\ref{line_free_energy}). First one can use the thermodynamic
definition of the compression modulus of
Eq.~(\ref{compression_modulus-definition}),
\begin{equation*}
c_{11}=a\partial^2_a\left[a\; f(a)\right],
\end{equation*}
and then gets via the relation of Eq.~(\ref{K}) the effective stiffness
\begin{equation}
\label{K-formula}
K=\frac{\pi}{16}\left(1+\frac{ag\Delta}{\pi^2 T^3}  \right)^{1/2}.
\end{equation}
Note that due to its linear dependence on density, the single line
free energy of Eq.~(\ref{line_free_energy}) does not contribute to
$K$.  The result of Eq.~(\ref{K-formula}) is compared in
Fig.~\ref{KvsTd-fig} to the numerical result as obtained from the
dimer model with isotropic disorder (see also
Ref.~[\onlinecite{Zeng+99}]) and from the dimer model with no random
energies on the horizontal bonds, i.e., $\epsilon_{ij\in h}=0$.  The
disorder strength $\Delta$ has been mapped here to $T_d$ according to
Eq.~(\ref{disorder_via_annealed}) and Eq.~(\ref{first_guess}),
respectively.  The agreement is very good over orders of magnitude
while the $\Delta$-$T_d$ relation from the comparison of annealed free
energy averages seems to fit better than the naive estimate. The
validity of the replica Bethe Ansatz calculation including its
sometimes debated analytical continuation in $n$ is nicely confirmed
by our comparison.  Interestingly, no deviation from the RBA result of
Eq.~(\ref{K-formula}) is found in the large disorder, i.e., low $T_d$
limit. In terms of the line lattice temperature $T$ modifications of
the result of Eq.~(\ref{line_free_energy}) must occur at $T \lesssim
T^*$, cf.~Eq.~(\ref{T_star}), since otherwise the free energy would
not converge to a finite ground state energy for $T \to 0$. However,
this problem no longer exists after the mapping to the dimer model
since the ratio $\sqrt\Delta/T$ of disorder and thermal energy is
controlled by the single parameter $T_d^{-1}$ in the dimer model. Thus
the pinning strength and thermal fluctuations cannot be varied
independently, and the crossover temperature $T^*$ can vanish.

\placefigure{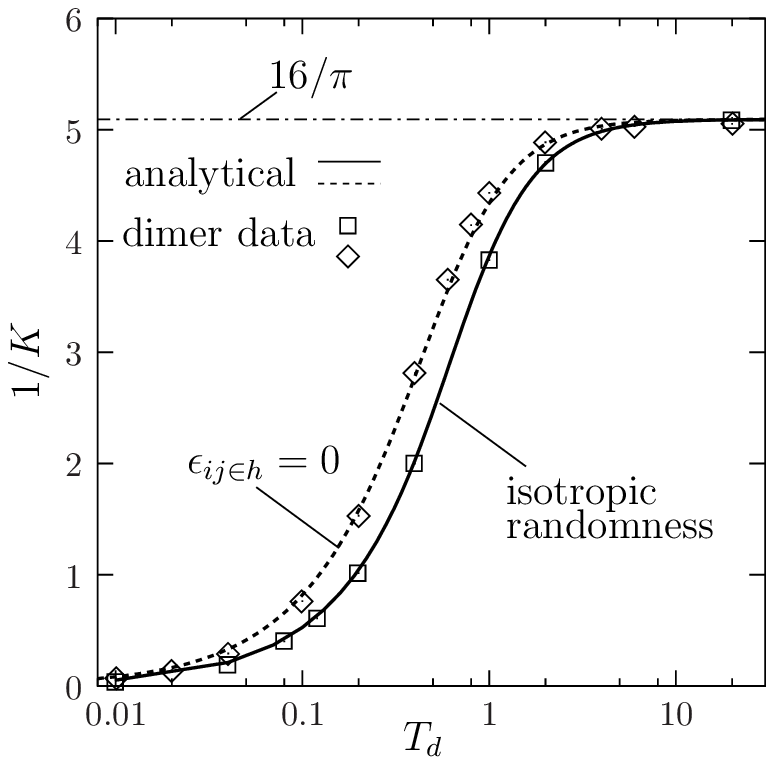}{0.9\linewidth}{Large scale stiffness $K$ as
  a function of the inverse disorder strength $T_d$.  Shown are the
  simulation results for the dimer model with isotropic random
  energies and for the model with random energies only on the
  vertical bonds together with the RBA result of
  Eq.~(\ref{K-formula}). In the isotropic case $T_d$ and $\Delta$ are
  related by Eq.~(\ref{disorder_via_annealed}) with $\xi_d=1.15$ while
  in the anisotropic case Eq.~(\ref{first_guess}) with $\xi_d=0.83$ has
  been used. }{KvsTd-fig}

In the pure limit $T_d\to\infty$ the stiffness approaches the value
$K=\pi/16$ in agreement both with the exact calculation in terms of
the nonrandom dimer model \cite{Henley97} and the mapping of the line
lattice without disorder to free fermions.  Hence the most accurate
simulation of a system of noncrossing lines by the dimer model is
demonstrated.  The precise value $K=\pi/16$ is of physical
significance as shown by the renormalization group (RG) scaling
dimension of disorder in RSG model Eq.~(\ref{H_RSG})
\begin{equation}
\lambda_\Delta=2(1-\pi/16\, K^{-1}),
\end{equation}
see also Ref.~[\onlinecite{Emig+03}].  The limiting value of
$K=\pi/16$ indicates that infinitesimal disorder is marginal and the
system is thus on the borderline between a glassy and a thermal free
phase.  Any finite amount of disorder increases $K$ which in turn
renders disorder a relevant perturbation, leading to a glassy phase.
This is consistent with the finding of Ref.~\onlinecite{Zeng+99} that
the correlation function of the height profile, i.e., the one of $h$
and not of $\delta h$, always indicates a low temperature glassy
behavior.

\subsection{Free energy, internal energy and entropy}
\label{free_energy}
%

We now come to a direct comparison of fundamental thermodynamic
quantities of the dimer model and of the continuum model for the line
lattice.  We start with the disorder averaged free energies
$\overline{F}=-T \overline{\ln \cZ}$.  Due to the different meanings
of temperature in the line and dimer context we will focus on the
logarithm of the partition functions $\overline{\ln \cZ}$.  When
relating the systems we remember the energy relation
Eq.~(\ref{energy_relation}) between dimer and line configurations.
Therefore, the partition functions $\cZ_d$ of the dimer model and
$\cZ_l$ of the line lattice are related by
\begin{eqnarray}
\label{E-ref-relation}
\ln \cZ_d +E_{\rm ref}/T_d = \ln \cZ_l
\end{eqnarray}
In the disorder average $\overline{\ln \cZ_d +E_{\rm ref}/T_d}$ the
reference energy $E_\text{ref}=H_\text{ref}(\{\epsilon'_{ij}\})$,
cf.~Eq.~(\ref{energy_relation}), does not contribute as the bond
energies are drawn from a Gaussian distribution with zero mean.
Higher moments of the free energy, however, contain contributions from
$E_{\rm ref}$. In the simulations of the dimer model the moments of
the density $\ln (\cZ_l)/(bL)^2$ are measured by taking into account
the reference energy $E_\text{ref}$.  The RBA, however, provides the
statistics of the free energy density of the line lattice
[Eq.~(\ref{line_free_energy})] which we denote by $f_l$ in the
following. Both quantities are related by
\begin{equation}
  \label{eq:dimer-ll-rel}
  \frac{\ln \cZ_l}{(bL)^2}=-\frac{A_l}{A_d} \frac{f_l}{T},
\end{equation}
where $A_l=L_x L_z$ and $A_d=(bL)^2$ are the system sizes of the line
lattice and dimer model, respectively. Here we have to pay attention
to the fact that the two models are only equivalent after a rescaling
of the $z$ coordinate as explained for the mapping between the models
D and E in Sec.~\ref{sec_mapping}.  According to
Eq.~(\ref{volume_relation}) we have
\begin{equation*}
\frac{A_l}{A_d} =
\frac{a}{16 K}\frac{g}{T}=
\frac{a}{\pi}\frac{g}{T}\left(1+\frac{ag\Delta}{\pi^2T^3}\right)^{-1/2}
\end{equation*}
which has to be used in Eq.~(\ref{eq:dimer-ll-rel}).  The disorder
independent part of the single line mean free energy $\overline{f}_0$
cannot be calculated unambiguously by the RBA.  We thus combine the
disorder independent contributions in $\overline{f}_l$ such that in
the pure limit ($T_d \to \infty$) the known free energy of the dimer
model is matched.  In this limit, the partition functions of the dimer
model just counts the number of complete dimer coverings of the square
lattice. This is a complex combinatorial problem as any flip of one
dimer may necessitate a cascade of flips throughout the system.
Nevertheless, the result is exactly known to be \cite{Kasteleyn61}
\begin{equation}
\label{eq:pure-Zd}
\ln \cZ_d |_{T_d\to\infty}=\frac{G}{\pi} L^2
\end{equation}
in the thermodynamic limit with Catalan's constant $G=0.915966$.
Next, we translate the line lattice parameters to the dimer model
along Table~\ref{mapping_table} and get for the dimer model without
horizontal energies, ($\epsilon_{ij\in h}=0$),
\begin{equation}
\label{f_fit_formula}
\frac{\overline{\ln \cZ_l}}{L^2}=\frac{1}{\sqrt{\pi^2+4\xi_d/(bT_d^2)}}
\left[ G+\frac{1}{T_d^2}\left(1-\frac{\xi_d}{2b}\right)-
\frac{1}{6}\frac{\xi_d^2}{b^2T_d^4}\right].
\end{equation}
If both vertical and horizontal bonds carry random energies one should
make the approximate replacement $1/T_d^2\to 2/T_d^2$, which can be
improved by the correction terms of Eq.~(\ref{disorder_via_annealed}),
as explained above.  The second term $\sim T_d^{-2}$ in the square
brackets of Eq.~(\ref{f_fit_formula}) comes from terms proportional to
disorder in both the single line free energy and the interaction part.
The last term $\sim T_d^{-4}$ comes from the term $\sim \Delta^2$ in
the single line free energy. When comparing the result of
Eq.~(\ref{f_fit_formula}) to the simulation results for the dimer
model, at first glance we find no agreement at all.  However, as we
will discuss shortly, there are indications that one might have to drop
the $T_d^{-4}$-term of Eq.~(\ref{f_fit_formula}). Doing so, we get the
plots of Fig.~\ref{f-fig} for the isotropic and the anisotropic random
bond energies.  Only the large $T_d$ limit was fixed by the known
result of Eq.~(\ref{eq:pure-Zd}), yet the agreement is excellent over
orders of magnitude down to small dimer temperatures. The only fitting
parameter $\xi_d/b$ arising from the disorder strength relation between
the discrete and the continuum model is found, as expected, to be of
order one, cf. the caption of Fig.~\ref{f-fig}.

\placefigure{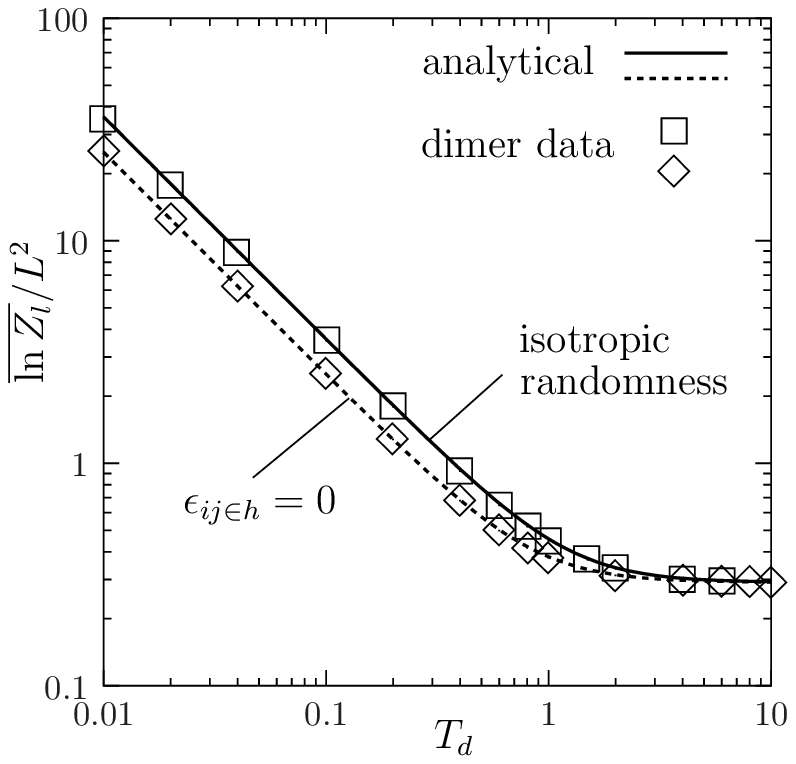}{0.9\linewidth}{Comparison of the
  disorder averaged free energy as obtained from the dimer model
  simulations and Eq.~(\ref{f_fit_formula}), respectively.
  $\overline{\ln Z_l}/L^2$ is plotted against dimer temperature
  $T_d\sim T/\sqrt{\Delta}$ for the isotropic dimer model
  ($\xi_d/b=0.96$) and the one with random energies only on the vertical
  bonds ($\xi_d/b=1.00$). The simulation data are for system size
  $L\times L=256 \times 256$.}{f-fig}

Why do we have to drop the $\Delta^2$-term of the single line free
energy [Eq.~(\ref{single_line_f})] to obtain agreement? One can, in
fact, imagine a number of reasons for this discrepancy between the RBA
result and the simulation data. The validity of the RBA itself has
been critically discussed, especially the interchange of thermodynamic
limit and replica number $n\to 0$ limit has been questioned. On the
other hand, the simulations are performed for discrete lattice
versions of the continuum model which has been solved by RBA.  We were
not able to find a conclusive answer to what causes the absence of the
$\Delta^2$-term in the simulation data. But in connection to this it
is interesting to remind of a numerical analysis of the average free
energy of a single directed polymer in a random potential via a
transfer matrix method in Ref.~\onlinecite{Krug+92}.  In
Fig.~\ref{Krug-plot} the simulation data of Table~II of
Ref.~\onlinecite{Krug+92} are plotted, giving the average free energy
as a function of disorder strength.  A plot of this kind had not been
shown in the cited reference.  However, it demonstrates that the data
obtained in Ref.~\onlinecite{Krug+92} agree with ours in {\em not}
finding support for the term $\sim \Delta^2$ in
Eq.~(\ref{single_line_f}).
\placefigure{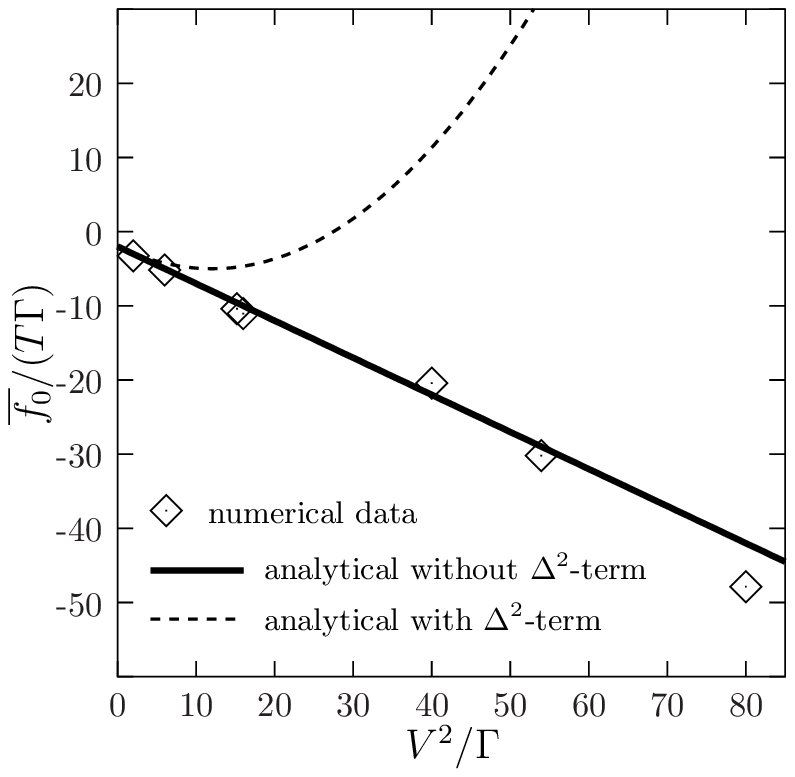}{0.85\linewidth}{Single line average free energy
  $\overline{f}_0$ plotted against disorder strength $V^2\sim \Delta$;
  data taken from Table~II of Ref.~[\onlinecite{Krug+92}], $\Gamma\sim
  1/g$. The solid and dashed curve shows the analytical result of
  Eq.~(\ref{single_line_f}) without and with the term $\sim \Delta^2$,
  respectively. The reason for the discrepancy is discussed in the
  text.}{Krug-plot}

Now we compare further thermodynamic quantities for the dimer model
and the line lattice.  The entropy and the internal energy of the
dimer model
\begin{eqnarray*}
\overline{S_d}&=&-\frac{\partial}{\partial T_d}\overline{F_d}=\frac{\partial}{\partial T_d}\left(T_d\overline{\ln \cZ_d}\right), \nonumber \\
\overline{U_d}&=&\overline{F_d}+T_d \overline{S_d}
\end{eqnarray*}
are easily be calculated from the RBA.  In the dimer model
simulations, the quenched averaged internal energy
$\overline{U_d}=\overline{\sum_{(ij)} p(ij)\epsilon_{ij}}$ with the
disorder configuration dependent dimer occupation probability $p(ij)$
of bond ($ij$) can be obtained quite easily since the polynomial
algorithm allows to calculate the probabilities $p(ij)$. The entropy
is then obtained from the free energy by subtraction.  A comparison of
the data for internal energy and entropy with the RBA prediction is
given in Fig.~\ref{su-fig} where we used the result of
Eq.~(\ref{f_fit_formula}) and $\overline{\ln \cZ_d}=\overline{\ln
  \cZ_l}$ to calculate $\overline{S_d}$ and $\overline{U_d}$.  We find
excellent agreement with $\xi_d/b\simeq1$.  The slope of
$\overline{S_d}$ at $T_d=0$ is calculated from
Eq.~(\ref{f_fit_formula}) to be
\begin{equation}
\frac{1}{L^2}\frac{\partial}{\partial T_d}\overline{S_d}|_{T_d=0}=
G\left(\frac{b}{\xi_d}\right)^{1/2}-
\frac{\pi^2}{8}\left(\frac{b}{\xi_d}\right)^{3/2} 
\left(1-\frac{\xi_d}{2b}\right)
\end{equation}
and matches the simulation data very well.

\placefigure{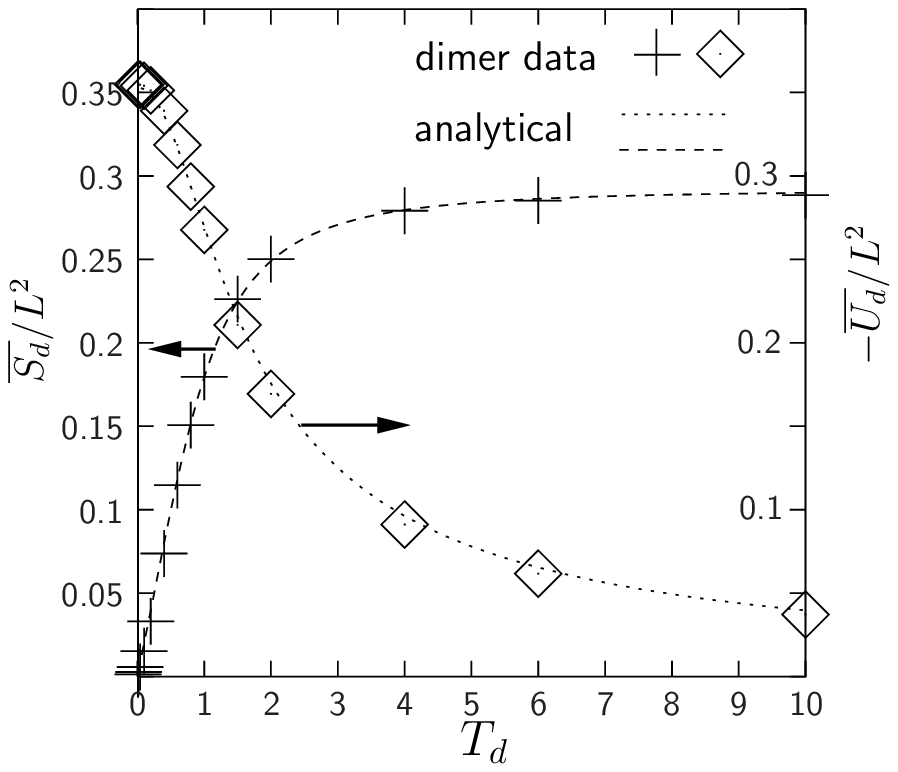}{0.98\linewidth}{Comparison between simulation
  data and RBA result for the quenched average entropy density
  $\overline{S_d}/L^2$ ($\xi_d/b=0.98$) and internal energy density
  $\overline{U_d}/L^2$ ($\xi_d/b=1.00$). The data are for system size
  $L=256$.}{su-fig}

Summarizing, the quantitative agreement between the RBA results for
the line lattice and simulation data for the dimer model is very
satisfying; it is even more surprising for the thermodynamic
potentials than for the large scale stiffness. The latter is expected
to show universality in the sense that it does not depend on
microscopic details of the model, while the former receive
contributions from all scales.  A priori, the sensitivity to the
contribution from modes close to the UV cutoff might have been
expected to be important.  However, our above results indicate that
the effect of small scales can be simply accounted for by the single
fit parameter $\xi_d/b$ which was found to be very close to the
naively expected value of one.

\subsection{Higher moments}
\label{higher_moments}

Higher cumulants of thermodynamic quantities describe sample-to-sample
fluctuations in experimental setups of mesoscopic dimensions while for
macroscopic systems their scaling will give information on the
selfaveraging behavior. Analytic expressions for higher cumulants of
the free energy are available \cite{Emig+01} only in the strong
disorder or dilute limit $l_d\rho \ll 1$, see
Eq.~(\ref{moments_perturbative}). At higher densities we have to
resort to a numerical computation of the ground state energy which
then yields $\ln \overline{\cZ^n}$ via Eq.~(\ref{Z_replica}), and thus
the cumulants of the free energy. In the following we are interested
in a polynomial expression for $E_0(n)$. Thus we have to expand in
Eq.~(\ref{E_0}) the integral of the kinetic energy with respect to
$n$. We introduce the dimensionless integral
\begin{eqnarray}
 \label{E_tilde}
 \tilde E_\text{kin}(n)&\equiv&
b^3 \int_{-K}^K \!dk\;k^2 \varrho(k)=(bK)^3\int_{-1}^1 \!dy \;y^2 
\tilde\varrho(y)\nonumber \\
&\equiv& \sum_{j\ge 0} \varepsilon_{j+1}(\rho l_d)\, n^j,
\end{eqnarray}
defining the expansion coefficients $\varepsilon_j(\rho l_d)$ which
depend only on the dimensionless parameter $\rho l_d$. With this
definition, explicit formulas for the variance (second cumulant) and
the skewness (third cumulant) of the reduced free energy can be
obtained from the RBA. Using Eqs.~(\ref{Z_replica}), (\ref{E_0})
together with Eq.~(\ref{E_tilde}) and the perturbative result of
Eq.~(\ref{moments_perturbative}) we get
\begin{widetext}
\begin{eqnarray}
  \frac{\overline{(\ln Z_l)^2_c}}{L^2}
  =\left\{
    \begin{array}{lcll}
      \displaystyle
      -\frac{A_l}{A_d}\frac{Tb^2}{g}\,\varepsilon_2(\rho l_d) & \to & 
\displaystyle -\frac{A_l}{A_d} \frac{1}{2} \,\varepsilon_2(T_d)&
      \quad\mbox{(exact RBA)}\\
      \displaystyle
      \frac{A_l}{A_d} \frac{4}{3\pi}\left(\frac{b\Delta}{a T^2}\right)^{3/2}
\left(\frac{bg}{T}\right)^{1/2} & \to & 
\displaystyle \frac{A_l}{A_d} \frac{2}{3\pi} \frac{(\xi_d/b)^{3/2}}{T_d^3} &
\quad\mbox{(perturbative, $T_d \ll 1$)},
    \end{array}
  \right.
\label{highermoments_expressions2}
\end{eqnarray}
\begin{eqnarray}
  \frac{\overline{(\ln Z_l)^3_c}}{L^2}
  =\left\{
    \begin{array}{lcll}
      \displaystyle
      \frac{A_l}{A_d}
\left\{ \left(\frac{\Delta}{T^2}\right)^2\frac{bg}{T} \frac{b}{4 a} 
-\frac{3 T b^2}{g}\, \varepsilon_3(\rho l_d)
\right\} & \to & \displaystyle \frac{A_l}{A_d}
\left\{ \frac{1}{4} \frac{(\xi_d/b)^2}{T_d^4} -\frac{3}{2}\, 
\varepsilon_3(T_d)\right\} &
\quad\mbox{(exact RBA)}\\
      \displaystyle \frac{A_l}{A_d}\left(\frac{3}{\pi^2}-\frac{1}{4}\right)
 \left(\frac{\Delta}{T^2}\right)^2
\frac{bg}{T}\frac{b}{a}
& \to & 
\displaystyle \frac{A_l}{A_d} \left(\frac{3}{\pi^2}-\frac{1}{4} \right)
\frac{(\xi_d/b)^2}{T_d^4} &
\quad\mbox{(perturbative, $T_d \ll 1$)}.
    \end{array}
  \right.
\label{highermoments_expressions3}
\end{eqnarray}
\end{widetext}
Here, the mapping $gb/T \to 2$, $\Delta/T^2\to (\xi_d/b)/T_d^2$, $a\to
2b$, $\rho l_d \to (b/\xi_d) T_d^2/4$ between the line lattice and the
dimer model with random energies only on the vertical bonds has been
applied.  Also the rescaling of the volume according to
Eq.~(\ref{volume_relation}) has to be applied for the comparison. In
terms of the dimer temperature $T_d$, the rescaling factor reads
\[
\frac{A_l}{A_d}=\frac{ag}{\pi T}\left(1+\frac{ag\Delta}{\pi^2T^3}\right)^{-1/2}
\to \frac{4}{\pi}{}\left(1+\frac{4\xi_d/b}{\pi^2 T_d^2}\right)^{-1/2}.
\]

In order to compare the above RBA results to simulation data for the
dimer model over the whole range of disorder strength, we solve the
Eqs.~(\ref{BA_equation1}),~(\ref{BA_equation2}) for $\rho(k)$
numerically.  From $\rho(k)$ the kinetic energy of Eq.~(\ref{E_tilde})
and thus the expansion coefficients $\varepsilon_j$ are obtained.  For
a numerical treatment it is useful to rewrite Eq.~(\ref{BA_equation1})
in the form
\begin{equation}
\label{BA_equation1b}
y=\frac{1}{n}\int_{-1}^1 \!dy'\;  g_n[K l_d (y-y')]\,\tilde\varrho(y')
\end{equation}
with $y=k/K$ and $\tilde\varrho(y)=\varrho(Ky)$. At fixed $K l_d$, the
dimensionless function $\tilde\varrho(y)$ is computed by the inversion
of the discretized integral equation. This inversion is quite delicate
since the present kind of inverse problem -- a Fredholm integral
equation of the first kind -- is extremely badly conditioned.  An
adequate treatment is, however, possible by use of, e.g., the method
of singular value decomposition (SVD) of the discretized integral
kernel \cite{numerical_recipes}.  The solution $\tilde\varrho(y)$ can
only be computed for a given value of $K l_d$.  With the so obtained
solution we can calculate the right hand side of
\begin{equation*}
\frac{\rho}{K}=\int_{-1}^1\!dy\; \tilde\varrho(y).
\end{equation*}
The dimer model density $\rho=1/(2b)$ then fixes $K$ and we thus
obtain the disorder strength $1/l_d$ that had been implied by our
initially chosen value for $K l_d$.  In this approach we cannot -- due
to the coupling of the BA equations -- modify $n$ and $l_d$
independently.  A modified $n$ implies a modified $K$, which results
in a different value of $l_d$. We hence have to adjust the parameter
$K l_d$ in Eq.~(\ref{BA_equation1b}) upon change of $n$ such that
$l_d$ remains constant. In practice, this is realized by the simple
method of nested intervals. The necessary number of discretization
points for the integral Eq.~(\ref{BA_equation1b}) depends crucially on
considered range of parameters.  An increase in the length scale $l_d$
stretches the integral kernel $g_n(k l_d)/n$ along the abscissa, a
decreasing $n$ does so along the ordinate.  Thus, in order to keep up
a given level of accuracy, the number of discretization points has to
increase like $l_d/n$.  We are unfortunately interested in small $n$
as we want to extract the coefficients $\varepsilon_j$ from the
behavior around $n=0$ and moreover in large $l_d$ as the result for
small $l_d$ is known analytically. Therefore it is important that the
reliability of the numerics can be checked in limiting cases where
analytical results for $\varrho(k)$ are available.  The limit of large
$l_d \to \infty$ or $n\to 1$ corresponds to lines without quenched
disorder which are described by free fermions with a constant density
$\varrho(k)=1/(2\pi)$.  In the inset of Fig.~\ref{rho-fig}, a plot of
the numerical solution in the latter case is displayed, showing very
good agreement with the analytical expectation.  An other limit which
can be compared to analytical results corresponds to $n\to 0$. Then
$\varrho(k)$ can be calculated perturbatively \cite{Emig+01} in $K
l_d$, see Eq.~(\ref{eq:rho-sol-n0}).  We compare the numerical
solution for $n=10^{-3}$ and $Kl_d=0.1$ with the result of
Eq.~(\ref{eq:rho-sol-n0}) in Fig.~\ref{rho-fig}. Again the comparison
is satisfactory while the mismatch at small $k/K$ is due to the
smallness of $n$ which necessitates a high discretization level. In
practice, $n$ is chosen as not to require more than $10^3$
discretization points for a relative accuracy of $10^{-4}$ in $\tilde
E_\text{kin}(n)$. Checks against the exact results for the quenched
averaged free energy [Eq.~(\ref{line_free_energy})] and the strong
disorder limit of the cumulants [Eq.~(\ref{moments_perturbative})] are
also satisfactory.
\placefigure{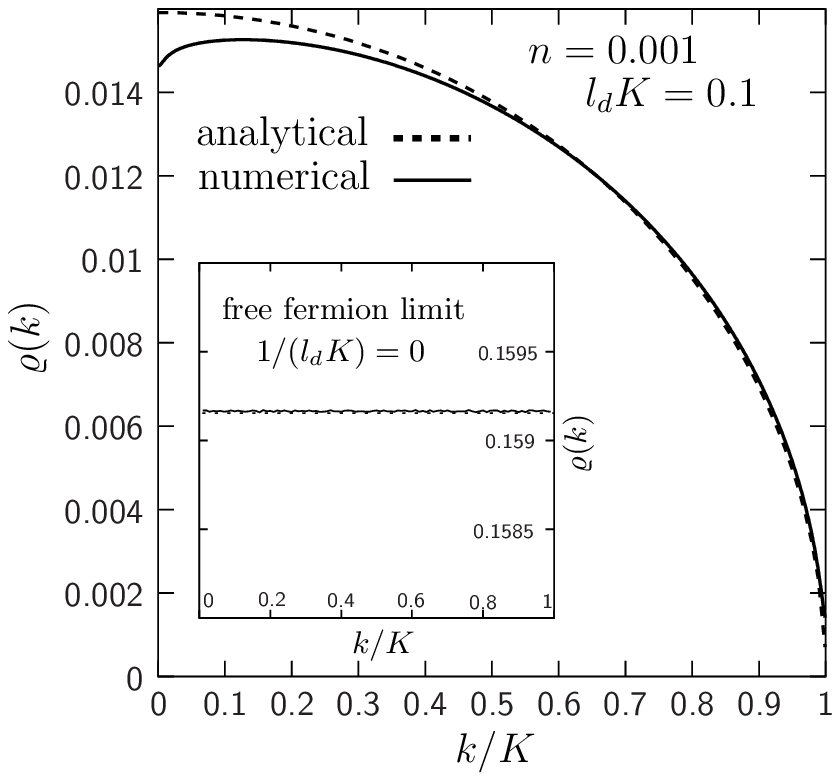}{0.8\linewidth}{Numerical solution
  $\varrho(k/K)$ for small $l_d K$, $n=0$ and for $l_d K \to \infty$
  (inset). The mismatch at small $k/K$ between the analytical
  expression and the numerical solution for $l_d K=0.1$ is reduced
  with an increasing number of discretization points. For the
  extraction of cumulants, larger values of $n/l_d$ have been used
  where a discretization level of $10^3$ points suffices.}{rho-fig}

Having determined the solution $\rho(k)$, the coefficients
$\varepsilon_j$ of Eq.~(\ref{E_tilde}) can be extracted from the
numerically calculated $\tilde E_\text{kin}(n)$ by repeated
extrapolation to $n=0$, subsequent subtraction of this value from the
finite-$n$ result and final division by $n$.  This straightforward
procedure is the best we could think of but it is still error-prone.
While the desire for a small extrapolation error requires having data
points as close as possible to $n=0$, a simple calculation of error
propagation shows that the error of a given data point at finite $n$
scales as $n^{-k}\delta \tilde E_\text{kin}$ where $k$ stands for the
order of the coefficient and $\delta \tilde E_\text{kin}$ for the
original error of the data.  However, we were able to achieve
sufficient high accuracy to extract reliable values for the second and
third cumulant.  The above described scheme was implemented with the
use of Numerical Recipe routines \cite{numerical_recipes}. The data
for the smallest disorder values needed the highest discretization
level and consumed about 200h computation time on a 2GHz processor for
the determination of the cumulants at a given value for $l_d$.

Before we compare the RBA predictions for the cumulants to the
simulation results for the dimer model we would like to make some
remarks on the computation of cumulants in the dimer model
simulations. In the replica theory the {\em cumulants} of random
thermodynamic quantities appear naturally, while in the simulations
the {\em moments} are immediately accessible.  Both are related as
follows.  If the generating functional of the moments $\{m_p\}$ is
\begin{equation*}
M(n)=1+nm_1+n^2\frac{m_2}{2!}+n^3\frac{m_3}{3!}+\dots,
\end{equation*}
then 
\begin{equation*}
\ln M(n)=n \kappa_1 +n^2\frac{\kappa_2}{2!}+n^3\frac{\kappa_3}{3!}+\dots
\end{equation*}
generates the cumulants $\{\kappa_p\}$.  The lowest cumulants
expressed in terms of moments are
\begin{eqnarray}
\label{kappa-m}
\kappa_1&=&m_1\nonumber\\
\kappa_2&=&m_2-m_1^2\nonumber\\
\kappa_3&=&m_3-3m_1m_2+2m_1^3.
\end{eqnarray}
From the dependence of Eq.~(\ref{Z_replica}) on the system size
$A_l=L_x\times L_z$ it follows immediately that all the free energy
cumulants scale linearly in $A_l$.  Apart from the average free
energy, the reduced cumulants
$\kappa_p/\kappa_1^p=\overline{F_c^p}/\overline{F}^p$ hence vanish in
the thermodynamic limit as $\overline{F_c^p}/\overline{F}^p\sim
A_l^{1-p}$ as one would expect from the central limit theorem.  The
distribution of the free energy becomes infinitely sharp in the limit
of large systems. In other words, the vortex-line array is self
averaging which in light of the infinite correlation length reflected
by logarithmic correlations had not been evident a priori.

For the determination of the cumulants from the simulation data the
following problem is entailed.  The moments $m_p$ scale with the
system size like $m_p\sim A_l^p$.  From Eq.~(\ref{kappa-m}) we see
that a cumulant of order $p$ has to be calculated as a sum of terms
that grow by a factor $A^{p-1}$ faster with the system size then the
cumulant itself.  Therefore, at a given accuracy of the simulation
data for the $m_p$, which primarily depends upon the number of
disorder samples, a limit is set to the system size up to where
cumulants can reliably be obtained.  This maximum system size
decreases with the order of the cumulant. On the other hand, finite
size effects have to be minimized as well and, as a consequence, at
the achieved precicion of $10^{-5}$ for the moments $m_p$ in the dimer
model, the variance can be trusted only for system up to size $L=64$
and the third cumulant up to size $L=16$.
  
In the following, we compare the simulation data for the cumulants
$\overline{(\ln Z_l)^p_c}/L^2$ for $p=2$, $3$ with the RBA predictions
of Eqs.~(\ref{highermoments_expressions2}),
(\ref{highermoments_expressions3}).  The variance $(p=2)$ as a
function of $T_d$ has been computed for the dimer model of size $L=64$
and is shown in Figs.~\ref{cumulant2-1-fig},~\ref{cumulant2-2-fig}
together with the RBA result.  With the only fitting parameter
$\xi_d/b= 0.8$ we find very good agreement. In addition, the plots
show that our numerical solution of the full Bethe ansatz equations
nicely confirms the perturbative solution at small $T_d$. The
deviation at larger $T_d$ shows that the numerical solution is
inevitable for a comparison with the simulation data. A closer look at
the analytical result of Eq.~(\ref{highermoments_expressions2}) and
the data points of Figs.~\ref{cumulant2-1-fig},~\ref{cumulant2-2-fig}
shows that $\overline{(\ln Z_l)^2_c}/L^2$ does {\em not} follow
strictly a power law in $T_d$.
\placefigure{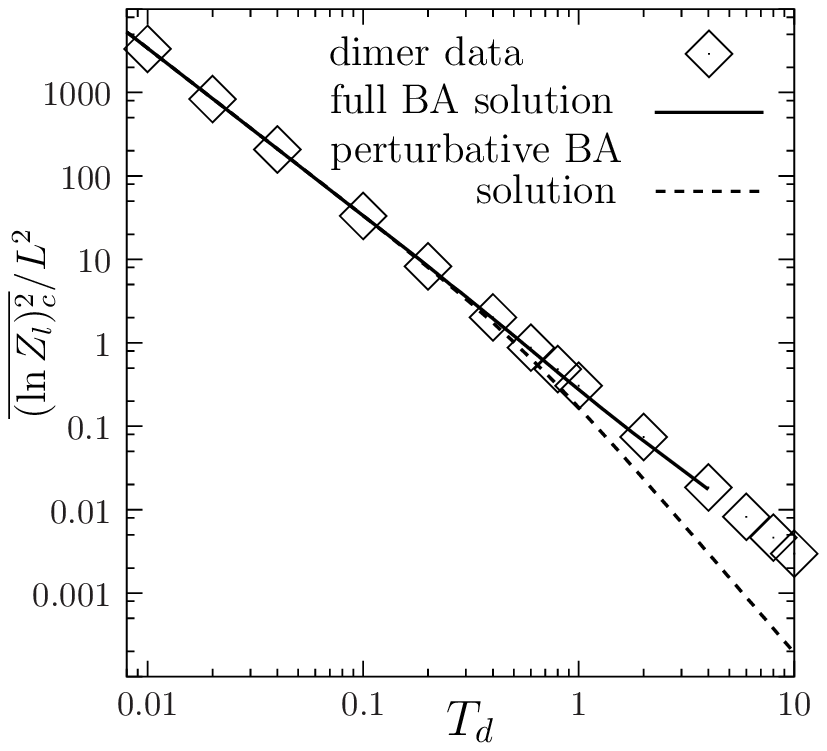}{0.9\linewidth}{Simulation data for the
  dimer model with random energies on the vertical bonds only
  (diamonds), the numerical result of the RBA equations (full line)
  and the perturbative RBA result (dashed line) for $\overline{(\ln
    Z_l)^2_c)}/L^2$. The fitting parameter is $\xi_d=0.8$, dimer
  system size is $L=64$.}{cumulant2-1-fig}
\placefigure{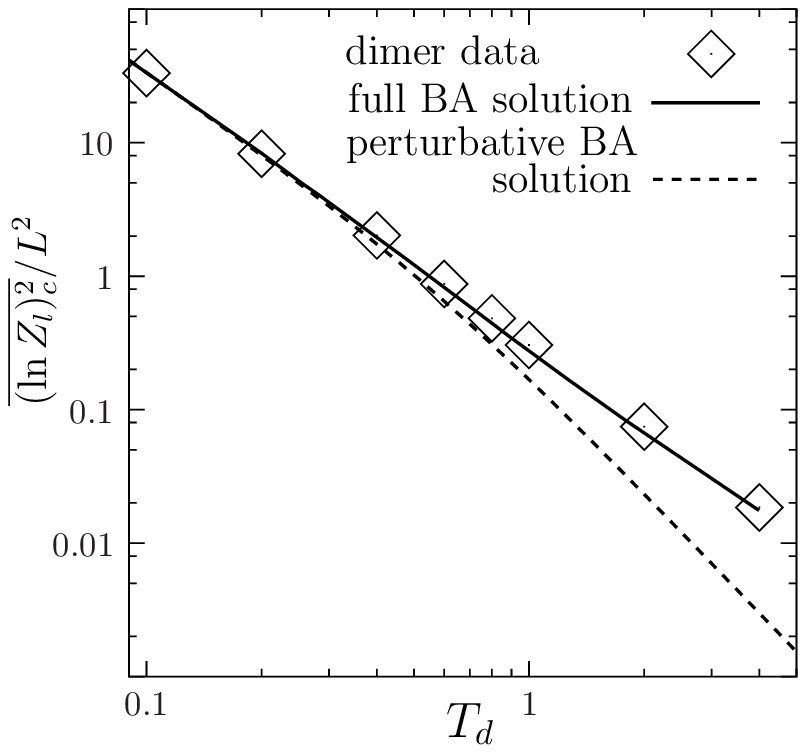}{0.9\linewidth}{Same as
  Fig.~\ref{cumulant2-1-fig} with a smaller range of
  $T_d$.}{cumulant2-2-fig}
The third cumulant (skewness, $p=3$) is shown in
Fig.~\ref{cumulant3-fig}.  For the reasons explained above, the dimer
data are to be trusted only for the small system $L=16$ and here only
for $T_d\lesssim 1$. In this range, no substantial deviations from the
perturbative evaluation of the cumulant, cf.
Eq.~(\ref{highermoments_expressions3}), is expected.  Indeed, the
agreement between theory and simulation is again very good with
$\xi_d/b=0.8$.  Although the numerically determined coefficient
$\varepsilon_3$ has an error of only $\sim 15\%$, the actual
uncertainty of the third cumulant is larger.  The reason is the
following. In the exact RBA result of
Eq.~(\ref{highermoments_expressions3}) the magnitude of the second
term $\sim \varepsilon_3$ amounts throughout the studied parameter
range to about $85\%$ of the first term.  Since the terms are
subtracted, the original error of $15\%$ gets amplified to $\sim
100\%$ in the final expression for the third cumulant. However, this
does not restrict the comparison since the perturbative RBA solution
could not have been corrected noticeably in the range $T_d\lesssim 1$
which is set by the simulation data at hand. Note that both the
simulation data and the perturbative RBA result consistently do not
obey a power law in $T_d$, see Fig.~\ref{cumulant3-fig}.

\placefigure{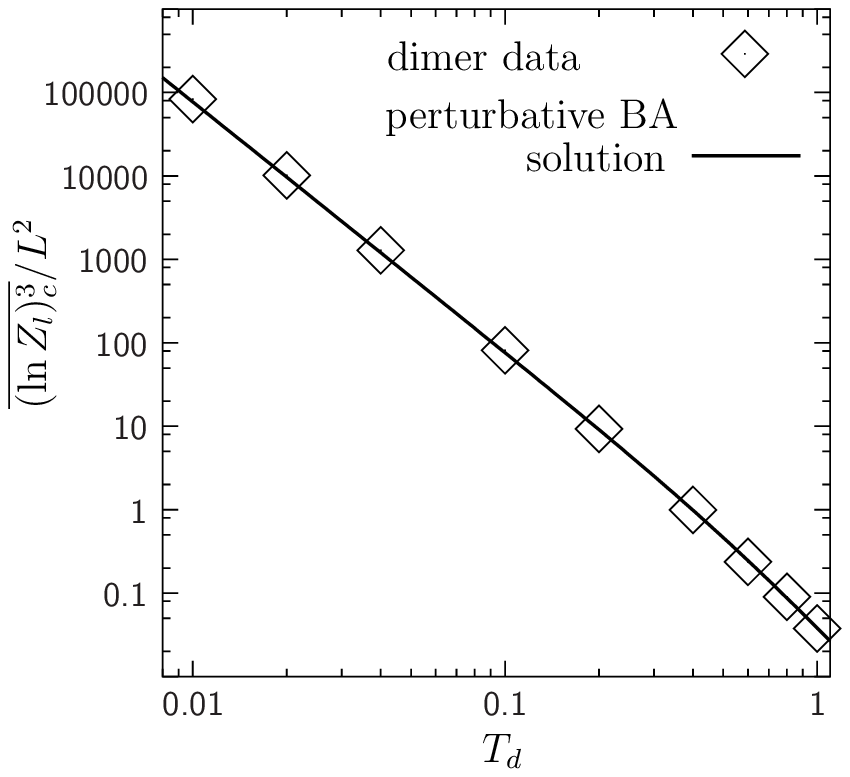}{0.9\linewidth}{Simulation data (diamonds) and
  perturbative RBA result (full line) for the third cumulant
  (skewness) $\overline{(\ln Z_l)^3_c}/L^2$. The dimer system size is
  $L=16$. Note that the data do not follow a power
  law.}{cumulant3-fig}

It must be noted that there is agreement for the third cumulant only
if the term $\sim n^3$ from the single line contribution $\sim
\frac{n}{12}(1-n^2)\Delta^2$ in Eq.~(\ref{moments_perturbative}) is
taken serious, otherwise not even the sign of the cumulant would
match. However, the part linear in $n$ stemming from the same
contribution we had to drop in the comparison of the average free
energy, being consistent with another independent study \cite{Krug+92}
of the single line free energy.  Since a single line (or directed
polymer) is expected to have an asymmetric disorder free energy
distribution, we do not argue for a complete irrelevance of the
contributions $\sim \Delta^2$ in the single line terms of
Eq.~(\ref{moments_perturbative}), consistent with our simulation data
for the third cumulant, but suggest a reassessment of the free energy
distribution of a single line in random media.

%
\subsection{Specific heat}
\label{response_functions}

The specific heat of disordered systems is strongly influenced by the
complex nature of thermal excitations about the pinned ground state.
In the context of spin glasses the excitations are considered as
droplets, i.e., connected regions in which the thermal activated
configurations differ from the ground state configuration
\cite{FisherDS+87,FisherDS+88}. Droplets appear on all length scales
with the lowest energy ones appearing on largest length scales. Here
we would like to test the RBA prediction for the mean specific heat by
comparing to our simulation results. The disorder averaged specific
heat of the dimer model with random energy on horizontal and vertical
bonds can be measured via the thermal fluctuations of the dimer
energy,
\begin{equation}
\label{eq:c-from-Hd}
\overline{c_d}=L^{-2}\overline{\frac{\langle H_d^2 \rangle - 
\langle H_d \rangle^2}{T_d^2}},
\end{equation}
where $H_d$ is the Hamiltonian of the dimer model, see
Sec.~\ref{sec:dimer_model}.  From the RBA result for the free energy
Eq.~(\ref{line_free_energy}) (again without the $\Delta^2$-term in the
single line free energy) the specific heat is calculated easily using
\begin{eqnarray}
\overline{c_d}=
\frac{T_d}{L^2}\frac{\partial^2}{\partial T_d^2} (T_d \, \overline{\ln \cZ_d}),
\end{eqnarray}
where $\overline{\ln Z_d}=\overline{\ln Z_l}$ is given by
Eq.~(\ref{eq:dimer-ll-rel}).  After the mapping $\Delta/T^2 \to
2(\xi_d/b)/T_d^2$, $bg/T \to 2$, $\rho\to 1/(2b)$, the mean specific
heat of the dimer model reads
\begin{widetext}
\begin{eqnarray}
\label{eq:RBA-c-result}
\overline{c_d}&=&
\frac{2\pi^2\left[2\pi^2-(\xi_d/b)(\pi^2+4G)\right]T_d^3-
8 \left[2\pi^2-(\xi_d/b)(\pi^2+16G)\right](\xi_d/b)\, T_d}
{\left[\pi^2T_d^2+8(\xi_d/b)\right]^{5/2}}.
\end{eqnarray}
\end{widetext}
In Fig.~\ref{c-fig} we compare this RBA result to the simulation data
which we obtained via Eq.~(\ref{eq:c-from-Hd}). We find rather nice
agreement over the entire range of $T_d$ with the choice
$\xi_d/b=0.98$ for the only fitting parameter, which is consistent
with our findings above. There are no further adjustable parameters in
the comparison shown in Fig.~\ref{c-fig}.  Also note that the dimer
specific heat probes for the agreement of the RBA and simulations
predominantly in the region around $T_d\simeq 1$, the drop to zero for
small and large $T_d$ being generic rather than specific. So it can be
considered complementary to the mean free energy which tested for
amplitude and exponent at small $T_d$ while the large $T_d$ (or pure)
limit was fixed to the exactly known result, see Section
\ref{free_energy}.
\placefigure{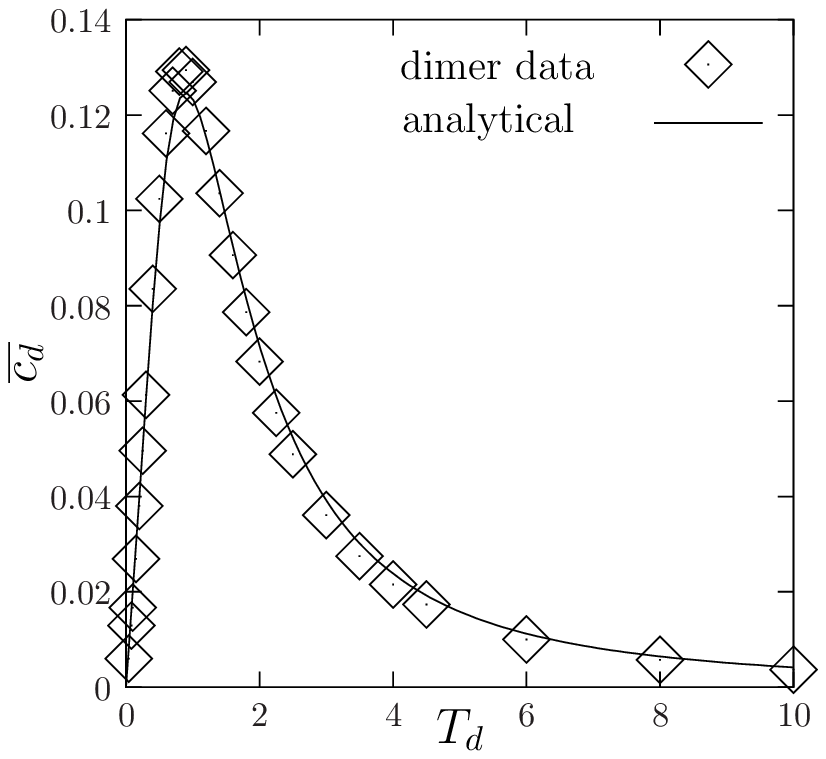}{0.9\linewidth}{Mean dimer specific heat
  $\overline{c_d}$. Shown are the simulation data (diamonds) and the
  analytical result of Eq.~(\ref{eq:RBA-c-result}) derived from the
  RBA (full line). The simulated dimer model has random energies on
  horizontal and vertical bonds, and is of size $L=256$.}{c-fig}

The linear low temperature behaviour of the specific is typical of
random systems. It can easily be understood by considering the lowest
lying excitations on a given length scale $\ell$.  On each scale for a
given disorder environment the ground state (with zero energy) and
the lowest excitation (with energy $E$) form a two level system,
whose specific heat is
\begin{equation*}
c_{\ell,E}=\frac{\partial}{\partial T} \frac{E e^{-E/T}}{e^{-E/T}+1}.
\end{equation*}
The excitation energies obey a length scale dependent disorder
distribution $p_\ell(E)$ and the contribution to the mean specific
heat from each scale is
\[
\delta \overline{c}_\ell =\int_0^\infty \!\!dE\; p_\ell(E)\, c_{\ell,E}
\sim p_\ell(0+)\, T.
\]
For a finite density of excitations at small energies, $\lim_{E\to 0+}
p_\ell(E)>0$, the specific heat as a superposition of exponentials
contributions from each scale will be {\em linear} at small
temperatures which is a famous insight of Anderson {\em et al.}
\cite{Anderson+72} Integration over all length scales allows to write
the mean specific heat as $\overline{c}=\int_0^L d\ell \, \delta
\overline{c}_\ell$ which becomes exact in the limit $T\to 0$.  In the
droplet theory of spin glasses the distribution function $p_\ell(E)$
is a central quantity.  From the finite size scaling of the mean
specific heat one can hope to obtain information on this distribution
function as with growing system size larger droplets will fit into the
system.  However, the droplets in our dimer model simulation seem to
be dominated by the system boundaries. The number of configurations of
lowest excitation energy that differ only on the boundaries from the
ground state configuration scales like the linear system size $L$,
yielding a $1/L$-{\em decay} for the specific heat (instead of a
growth from bulk droplets) to its asymptotic value .  In fact, this
scaling behavior is observed for the specific heat data of our dimer
model simulation, see Fig.\ref{c-slope-fig}.  Indeed, low-lying
excitations on the boundary can be easily identified. Consider a bond
on the boundary that is occupied by a dimer in the ground state.  A
configuration that does not cover this very bond may remain unchanged
on all the other bonds since simulations are done with open boundary
conditions. The missing energy on the bond is the excitation energy,
whose probability distribution, however, is nontrivial. The knowledge
of this distribution would allow to calculate the finite size scaling
of the slope of the specific heat at $T_d=0$. Due to the simplicity of
the boundary droplets the distribution could be computed with the
dimer algorithm.  The conditional probability that a bond is occupied
given its random energy is just the probability $p(ij)$ introduced
above in the calculation of the internal energy, see
Sec.~\ref{free_energy}. It can be easily obtained from the dimer
algorithm. The distribution of the boundary droplet energies is then
given by $p(ij) p(occ)/p(E)$ with $p({\rm occ})$, $p(E)$ being the
probabilities for the occupation and energy $\epsilon_{ij}=E$ of a
bond, respectively.

\placefigure{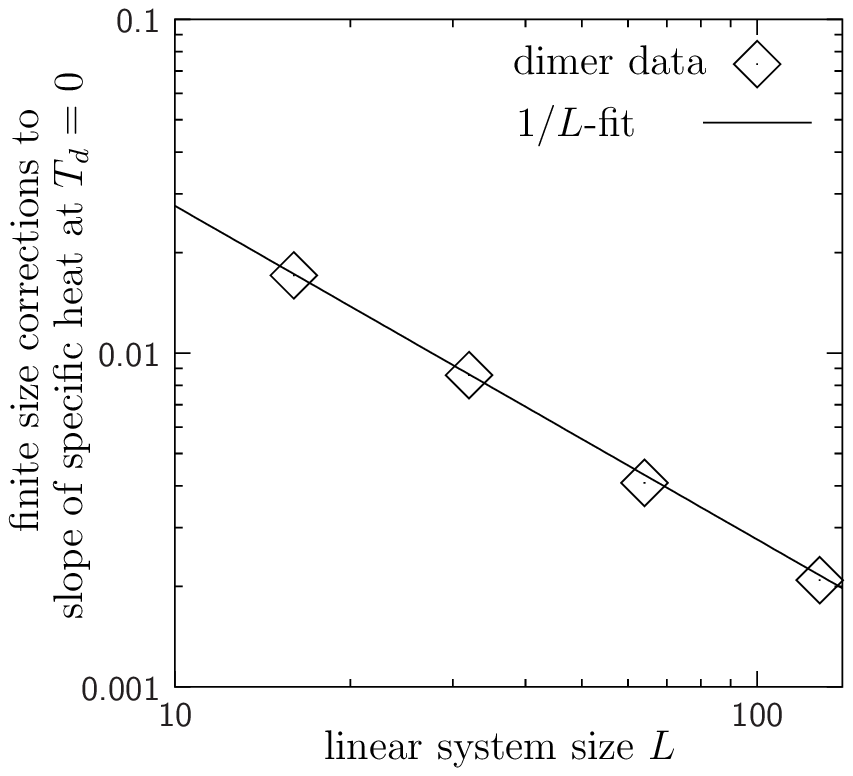}{0.9\linewidth}{Scaling of the slope of the
  dimer specific heat $\overline{c_d}$ at $T_d=0$ with the system
  size.}{c-slope-fig}

The smallest droplet excitation in the bulk is likewise easily
identified as the rotation of a plaquette that consists of two
opposite dimers. The probability distribution for the energy
difference of the two configurations is, however, not easily obtained.
It is complicated by the condition that the two dimers before the flip
must be part of the groundstate configuration. In the simulations, it
should be stressed, statistics of droplet energies can in principle be
measured systematically in the following straightforward procedure.
For a given disorder configuration the groundstate dimer covering is
determined.  Then the energy of one arbitrary {\em occupied} bond in
the bulk is set to infinity and the new groundstate is determined. It
will not contain the bond with infinite energy and hence have higher
energy than the original groundstate. The energy difference $E$
together with the diameter $\ell$ of the non-overlapping region of the
two ground states is measured. The statistics of these pairs of values
for many disorder realizations yield the distribution $p_\ell(E)$.
Quantitative support of the scaling prediction of droplet theory is in
reach considering the orders of magnitude over which the dimer model
can be simulated.

\section{Summary and Discussion}
\label{sec:discussion}

In this article we have compared recent exact replica Bethe ansatz
results for the planar line lattice with numerical simulations of the
classical random bond dimer model.  We found excellent agreement for a
large set of disorder averaged thermodynamic quantities, namely the
effective disorder renormalized elastic stiffness, the free energy,
the internal energy, entropy and the specific heat of the dimer model.
Characteristics of the disorder distribution of thermodynamic
quantities can also be obtained both from the Bethe ansatz and from
simulations and their agreement has explicitly been shown for the
variance and the skewness (third cumulant) of the free energy.  The
comparison thus confirms the Replica Bethe Ansatz calculation of
Ref.~\onlinecite{Emig+01} and makes the $(1+1)$-dimensional line
lattice one of the few glassy systems for which the validity of a
replica approach (without replica symmetry breaking) can be critically
tested in detail.  In the comparison, only one free parameter has been
used which is the ratio of the short scale regularization lengths of
the continuous line model and the lattice constant of the discrete
dimer model. The ratio is found to be consistently of the order of
unity for all studied thermodynamic quantities.  One term in the
single line free energy, first given in Ref.~\onlinecite{Kardar+85},
could not be confirmed by the simulations, in consistency with
numerical data of Ref.~\onlinecite{Krug+92}.  It would be interesting
to relate the droplet excitations of the dimer or line system to the
excitations of the related $SU(n)$ fermi gas in the limit $n\to 0$.
In view of the manifold links of dimer covering models to condensed
matter systems, especially of spin systems
\cite{FisherME66,Kac+52,Moessner+02} the link to the exactly solved
disordered line lattice model might prove useful in future
applications.

\acknowledgments

Interesting discussions with M.~Kardar and S.~Scheidl are gratefully
acknowledged. This research was supported by the Deutsche
Forschungsgemeinschaft (DFG) through an Emmy Noether grant (S.B. and
T.E.) and by NSF Grant DMR-0049176 (A.T. and C.Z.).

\end{document}